\documentclass[12pt]{article}      
\usepackage{array}      
\usepackage{epsfig}      
\usepackage{amssymb}      
\usepackage{graphics,graphpap}      
\usepackage{amssymb}      
\usepackage{amsmath}      
\usepackage[usenames]{color}     
\usepackage{slashed}      
\usepackage{graphicx}				     
      
\renewcommand{\thefootnote}{\fnsymbol{footnote}}      
\def \cez {{c_{12}}}      
\def \ced {{c_{13}}}      
\def \cev {{c_{14}}}      
\def \czd {{c_{23}}}      
\def \czv {{c_{24}}}      
\def \cdv {{c_{34}}}      
\def \sez {{s_{12}}}      
\def \sed {{s_{13}}}      
\def \sev {{s_{14}}}      
\def \szd {{s_{23}}}      
\def \szv {{s_{24}}}      
\def \sdv {{s_{34}}}      
\def \ded {{\delta_{13}}}      
\def \dev {{\delta_{14}}}      
\def \dzv {{\delta_{24}}}      
\def \dzv {{\delta_{24}}}      
      
\newcommand{\nn}{\nonumber}

\def\s#1{\setbox0=\hbox{$#1$}%
  \rlap{\ifdim\wd0>.7em\kern.22\wd0\else\kern.1\wd0\fi /}#1}

      
\begin{document}      
      
\begin{titlepage}      
\begin{flushright}      
\begin{tabular}{l}      
 DO-TH 10/09
 \\   
 TTK-10-32
 \\
 SFP/CPP-10-35      
\end{tabular}      
\end{flushright}      
\vskip1.5cm      
\begin{center}      
{\Large \bf \boldmath      
Less space for a new family  of fermions  
}      
\vskip1.3cm       
{\sc      
Otto Eberhardt     \footnote{Otto.Eberhardt@physik.uni-regensburg.de}$^{,1}$,      
Alexander Lenz      \footnote{Alexander.Lenz@physik.uni-regensburg.de}$^{,2,1}$,      
\\ and \\      
J{\"u}rgen Rohrwild \footnote{rohrwild@physik.rwth-aachen.de}$^{,3}$}      
\vskip0.5cm      
$^1$ Institut f\"ur Theoretische Physik, \\ Universit\"at Regensburg,  D--93040 Regensburg, Germany      
\vskip0.5cm      
$^2$ Institut f\"ur Physik, \\ Technische Universit\"at Dortmund,   D--44221 Dortmund, Germany      
\vskip0.5cm      
$^3$ Institut f\"ur Theoretische Teilchenphysik und Kosmologie, \\ RWTH Aachen University,      
     D--52056 Aachen, Germany

\vskip1cm

{\large\bf Abstract\\[10pt]} \parbox[t]{\textwidth}{  
We investigate the experimentally allowed parameter space of an extension of the standard model (SM3) 
by one additional family of fermions. Therefore we extend our previous study of the CKM like  
mixing constraints of a fourth generation of quarks.   
In addition to the bounds from tree-level determinations of the 3$\times$3 CKM elements and       
FCNC processes ($K$-, $D$-, $B_d$-, $B_s$-mixing and the decay $b \to s \gamma$)      
we also investigate the electroweak $S$, $T$, $U$ parameters, the angle $\gamma$ of the unitarity triangle   
and the rare decay $B_s \to \mu^+ \mu^-$. 
Moreover we improve our treatment of the QCD corrections compared to our previous analysis.   
We also take leptonic contributions into account, but we neglect the mixing among leptons.   
As a result we find that typically small mixing with the fourth family is favored, but still 
some sizeable deviations from the SM3 results are not yet excluded. 
The minimal possible value of $V_{tb}$ is 0.93. Also very large CP-violating effects in  
$B_s$ mixing seem to be impossible within an extension of the SM3 that consists of an  
additional fermion family alone.  
We find a delicate interplay of electroweak and flavor observables,   
which strongly suggests that a separate treatment of the two  sectors is not feasible.   
In particular we show that the inclusion of the full CKM dependence of the $S$ and $T$ parameters in principle allows the existence of a degenerate fourth generation
of quarks. 
}     
      
\vfill

\end{center}      
\end{titlepage}      
      
\setcounter{footnote}{0}      
\renewcommand{\thefootnote}{\arabic{footnote}}      
\renewcommand{\theequation}{\arabic{section}.\arabic{equation}}      
      
\newpage      
      
\section{Introduction}      
\setcounter{equation}{0}      
Increasing the number of fermion generations    
(see \cite{Frampton:1999xi} for a review and \cite{Holdom:2009rf} for an update)    
is probably the most obvious extension of the usual standard model with three generations (SM3).    
Although being popular    
in the 80s, such a possibility was  discarded for a long time. Recently these models (SM4)   
celebrated a kind of resurrection. Partly, this was due to the fact that a fourth generation   
is not necessarily in conflict with electroweak precision observables   
\cite{Holdom:1996bn,Maltoni:1999ta,He:2001tp,Novikov:2001md,Novikov:2002tk,Yanir:2002cq,Frere:2004rh,Alwall:2006bx,Kribs:2007nz,Novikov:2009kc,Vysotsky:2009if,Erler:2010sk}.   
\\   
Besides being a straightforward extension of the SM3, an increase of the number of fermion   
generations leads also to several desired effects:   
\begin{itemize}   
\item  The authors of \cite{Novikov:2002tk,Frere:2004rh,   
       Kribs:2007nz,Novikov:2009kc,Vysotsky:2009if,Hashimoto:2010at} have shown that    
       a fourth generation softens the current low Higgs mass bounds from electroweak precision observables,   
       see e.g. \cite{Flacher:2008zq}, by allowing considerably higher values for the Higgs mass.     
\item It might solve problems related to baryogenesis:      
      An additional particle family could lead to a sizeable increase of the measure of    
      CP-violation, see \cite{Hou:2008xd,Hou:2010wf}.  
      Moreover, such an extension of the SM would increase the strength of the phase transition, see    
      \cite{Carena:2004ha,Fok:2008yg,Kikukawa:2009mu}.      
\item The gauge couplings  can in principle be unified without invoking SUSY \cite{Hung:1997zj}.      
\item New heavy fermions lead to new interesting effects due to their large Yukawa couplings, see e.g   
      \cite{Hung:2009hy,Hung:2009ia}. Moreover dynamical electroweak symmetry breaking might be    
      triggered by these heavy new  fermions \cite{Holdom:1986rn,Carpenter:1989ij,King:1990he,Hill:1990ge,   
      Hung:1996gj,Holdom:2006mr,Burdman:2008qh,Hashimoto:2009ty,Antipin:2010it,Delepine:2010vw,Simonov:2010wd}.   
      This mechanism can also be incorporated    
      in models with warped extra dimensions, as done in   
      \cite{Burdman:2007sx,Burdman:2009ih}.   
\end{itemize}   
There are also some modest experimental deviations that could be explained by the existence of a fourth generation:   
\begin{itemize}   
\item A new family might cure certain problems in flavor physics (CP-violation in $B_s$-mixing, $K-\pi$-puzzle,    
    $\epsilon_k$ anomaly,...) see e.g.      
      \cite{Soni:2009fg,Soni:2008bc,Hou:2006mx,Arhrib:2006pm,Hou:2005hd} for some recent work and     
      e.g. \cite{Hou:1986ug,Hou:1987vd} for some early work on 4th generation effects on     
      flavor physics.    
\item Investigations of lepton universality show a value of the PMNS element $V_{e4} \neq 0$   
      at the 2.5 $\sigma$ level \cite{Lacker:2010zz}.  
\end{itemize}   
For more arguments in favor of a fourth generation see e.g.  \cite{Holdom:2009rf} and also
\cite{Bracic:2005ic,Bennett:2007hf,Belotsky:2008se,Borstnik:2010xs}. 
We conclude the list    
by repeating our statement from \cite{Bobrowski:2009ng}:   
{\it In view of the (re)start of the LHC,    
it is important not to exclude any  possibility for new physics scenarios simply due to prejudices.}     
Direct search strategies for heavy quarks at the LHC are worked out e.g. in   
\cite{Burdman:2008qh,Arik:1996qd,Skiba:2007fw,Holdom:2007nw,Holdom:2007ap,Ozcan:2008qk,AguilarSaavedra:2009es,Holdom:2010fr}.  
Signatures and consequences for collider physics, such as the modification of production rates, have been studied e.g.~in  
\cite{Barger:1986jt,Burdman:2008qh,Chen:2009zzu,Bernreuther:2010uw,Liu:2010ze}. 
  
In this work we extend our analysis in \cite{Bobrowski:2009ng},    
where we performed an exploratory study of the    
allowed parameter range  for the CKM like mixing of hypothetical quarks of a fourth generation.   
Adding one generation of quarks results in several new parameters. In particular, we have the new masses   
$m_{b'}$ and $m_{t'}$, and nine parameters (six angles and three phases) in the $4 \times 4$ CKM matrix   
(compared to three angles and one phase in the SM3).  
Following our previous strategy, we consecutively add bounds on the   
CKM structure of the SM4 and perform a scan though the parameter space   
of the model to identify the allowed regions; while this treatment   
is insufficient to fit for the central values or standard deviation   
of the model parameters, it gives a very reasonable idea of  the  
experimentally possible parameter space allowing for statements on the size of effects  
of the model on particularly interesting flavor observables.  
Apart from the unitarity of the $4\times 4$ matrix and the direct bounds   
on the quark masses, the most important   
input comes from direct measurements of the absolute values of   
CKM matrix elements, e.g.~from $\beta$ decay. In \cite{Bobrowski:2009ng}  
the next step was the inclusion of flavor observables  
sensitive to FCNC, mediated i.e.~by box or penguin diagrams. This led to   
some surprising results regarding the possible size of the quark mixing  
with the fourth generation quark, as rather large values for the mixing   
angle $s_{34}$ could not be excluded.  
 
However, in \cite{Chanowitz:2009mz} Chanowitz found that the   
parameter sets, which we gave as an example for   
large mixing with the fourth generation, are excluded by electroweak precision constraints, in particular by the    
oblique corrections \cite{Peskin:1991sw}. Moreover, Chanowitz performed the whole electroweak fit for four   
different values of the mass of the $t'$ quark. Here, some assumptions were used:   
i) the lepton masses are fixed to $m_{l4} = 145$ GeV and $m_{\nu_4} = 100$ GeV,   
ii) lepton mixing is not included,   
iii) the mass difference of the heavy quark doublet is fixed to $m_{t'} - m_{b'} = 55$ GeV,   
iv) only mixing between the third and fourth family was included.  
Assumptions iii) and iv) were also tested in  \cite{Chanowitz:2009mz}.  
 
Therefore, we supplement the analysis of the flavor sector by the $S$, $T$ and $U$ parameters;   
also the lepton masses of the fourth generation have to be taken into  account.   
For the present work we assume that the neutrinos have Dirac character and   
neglect the possible mixing of the fourth neutrino in the lepton sector.  
Moreover, we extend the  set of our FCNC observables to include also $B_s \to \mu^+ \mu^-$ 
and we improve the simplified treatment   
of the decay $b \to s \gamma$ by using the full leading logarithmic result.    
Concerning the tree-level determination of the CKM elements we include now also the experimental   
results for the angle $\gamma$ of the unitarity triangle, which gives a direct constraint on CKM phases.     
Similar studies have been recently performed e.g. in   
\cite{Bashiry:2009zz,Eilam:2009hz,Soni:2010xh,Buras:2010pi}   
\\   
In Section 2 we present all experimental constraints we use in our analysis.   
We start with the parameterization of $V_{CKM4}$ in Section 2.1,   
next we discuss briefly tree-level determinations of CKM elements and direct mass limits.   
The electroweak parameters $S$, $T$ and $U$ will be   
investigated before reviewing the FCNC constraints.   
We end Section 2 with the allowed regions for deviations of the SM4 results from the SM3 values.   
\\   
In Section 3 we determine the bounds on the parameters of the model. After   
explaining our general strategy in 3.1, we present the results for the different mixing   
angles of $V_{CKM4}$, the new results for $V_{cx}$ and $V_{tx}$ ($x = d,s,b$), and allowed   
effects of a fourth generation in neutral meson mixing.   
\\   
In Section 4 we give a Wolfenstein-like expansion of the $4\times 4$ CKM matrix. 
With the additional information from the electroweak sector, tighter constraints on 
the fourth generation quark mixing can be utilized leading to a simplified expansion.  
\\   
We conclude with Section 5.  

\section{Constraints on $V_{CKM4}$}      
\setcounter{equation}{0}      
\subsection{Parameterization of $V_{CKM4}$}      
In the SM3 the mixing between  quarks is described by the unitary $3 \times 3$ CKM matrix    
\cite{Cabibbo:1963yz, Kobayashi:1973fv},     
which can be parameterized by three angles, $\theta_{12}$,  $\theta_{13}$ and $\theta_{23}$     
($\theta_{ij}$ describes the strength of the mixing between the $i$th and $j$th family) and     
the CP-violating phase  $\delta_{13}$. The so-called standard parameterization of $V_{CKM3}$      
reads      
\begin{equation}     
\label{CKM3}  
                   V_{CKM3} =      
\left( \begin{array}{ccc} c_{12} c_{13}  & s_{12} c_{13} & s_{13} e^{-i \delta_{13}} \\      
  -s_{12} c_{23} - c_{12} s_{23} s_{13} e^{i \delta_{13}} &  c_{12} c_{23} - s_{12} s_{23} s_{13} e^{i \delta_{13}} & s_{23} c_{13} \\     
   s_{12} s_{23} - c_{12} c_{23} s_{13} e^{i \delta_{13}} & -c_{12} s_{23} - s_{12} c_{23} s_{13} e^{i \delta_{13}} & c_{23} c_{13} \\     
       \end{array}      
\right)     
                   \end{equation}     
with     
\begin{equation}     
s_{ij} := \sin (\theta_{ij}) \, \, \, \, \, \mbox{and} \, \, \, \, \,     
c_{ij} := \cos (\theta_{ij})\, .     
\end{equation}     
Extending the minimal standard model to include a fourth family of fermions (SM4) introduces    
3 additional angles in the CKM matrix $\theta_{14}, \theta_{24}$  and $\theta_{34}$   
and 2 additional CP-violating phases $\delta_{14}$ and $\delta_{24}$.      
To determine the allowed range for these new parameters we use an exact parameterization    
of the $4 \times 4$ CKM matrix. We have chosen the one suggested by    
Botella and Chau \cite{Botella:1985gb}\footnote{In the published paper       
of Botella and Chau there is a typo in the element $V_{td}$: in the last term of $V_{td}$ the factor $s_y$    
has to be replaced by $c_y$.},   
Fritzsch and Plankl \cite{Fritzsch:1986gv}\footnote{In the published paper       
of Fritzsch and Plankl there is a typo in the element $V_{cb}$: the factor $c_{23}$    
has to be replaced by the factor $s_{23}$.}      
and also by Harari and Leurer \cite{Harari:1986xf}.

{\footnotesize      
\begin{equation}     
\hspace{-1cm}      
 V_{CKM4} = \left( \begin{array}{cccc}      
\cez \ced \cev  & \ced \cev \sez & \cev \sed e^{-i\ded} & \sev e^{-i\dev} \\      
& & &\\      
-\czd \czv \sez -\cez \czv \sed \szd e^{i\ded} &\cez \czd \czv -\czv \sez \sed \szd e^{i\ded} &      
      \ced \czv \szd & \cev \szv e^{-i\dzv} \\      
-\cez \ced \sev \szv e^{i(\dev-\dzv)} & -\ced \sez \sev \szv e^{i(\dev-\dzv)} &      
      -\sed \sev \szv e^{-i(\ded+\dzv-\dev)} &  \\      
& & & \\      
-\cez \czd \cdv\sed e^{i\ded} +\cdv \sez \szd & -\cez \cdv \szd - \czd \cdv \sez \sed e^{i\ded} &      
      \ced\czd\cdv & \cev \czv \sdv  \\      
-\cez \ced \czv \sev \sdv e^{i \dev} & -\cez \czd \szv\sdv e^{i\dzv} & -\ced\szd\szv\sdv e^{i\dzv} &      
 \\      
+\czd\sez\szv\sdv e^{i\dzv}& -\ced \czv \sez\sev\sdv e^{i\dev} & -\czv\sed\sev\sdv e^{i(\dev-\ded)} &      
      \\      
+\cez \sed \szd\szv\sdv e^{i(\ded+\dzv)} & +\sez \sed \szd \szv \sdv e^{i(\ded+\dzv)} & & \\      
& & &\\      
-\cez \ced \czv \cdv \sev e^{i\dev } & -\cez \czd \cdv \szv e^{i\dzv}+\cez \szd \sdv &      
       -\ced \czd \sdv & \cev \czv \cdv\\      
+\cez \czd \sed \sdv e^{i \ded} & -\ced \czv \cdv \sez \sev e^{i \dev}&      
       -\ced \cdv \szd \szv e^{i\dzv} & \\      
+\czd \cdv \sez \szv e^{i\dzv}-\sez \szd \sdv & +\czd \sez \sed \sdv e^{i \ded} &      
       -\czv \cdv \sed \sev e^{i(\dev-\ded)} & \\      
+\cez \cdv \sed \szd \szv e^{i(\ded+\dzv)} & +\cdv \sez  \sed \szd \szv e^{i(\ded + \dzv)} & &      
\end{array} \right) \label{eq:CKM4FP}     
\end{equation}      
}      
For our strategy the explicit form of $V_{CKM4}$ does not matter, it is only important    
that the parameterization is exact.   
Besides the nine parameters of $V_{CKM4}$ we have also the masses of the fourth generation particles, which we denote as $m_{b'}, m_{t'}, m_{l_4}$ and $m_{\nu_4}$. We do not include leptonic mixing, yet.

\subsection{Experimental bounds}     
In this section we summarize the experimental constraints that have to be    
fulfilled by the parameters of the fourth family.   
\\   
The elements of the 3$\times$3 CKM matrix have been studied      
intensely for many years and precision data on most of them      
is available. In principle there are two different ways to     
determine the CKM elements. On the one hand, they    
enter charged weak decays already at tree-level and a measurement     
of e.g. the corresponding decay rate provides direct information on     
the  CKM elements (see e.g. \cite{Battaglia:2003in} and references therein).     
We will refer to such constraints as {\it      
tree-level constraints}. On the other hand, processes involving     
 flavor-changing neutral currents (FCNC) are forbidden at tree-level     
and only come into play at loop level via the renowned Penguin and     
Box diagrams. These processes provide strong bounds ---      
referred to as {\it FCNC constraints} --- on the structure of the      
CKM matrix and its elements as well as on the masses of the heavy    
virtual particles appearing in the loops.      
\\   
We will start with the tree-level constraints, since they only   
depend on the CKM elements and not on the fermion masses. Next we consider    
mass constraints on the fourth family members from direct searches at colliders.   
Since the oblique electroweak parameters are expected to reduce the allowed range of masses   
for a new fermion family notably, we consider them next and finally we discuss the FCNC constraints.   
%
%
%
%
%
%
%
\subsubsection{Tree-level constraints for the CKM parameters}     
\label{sec:tree}  
Since the (absolute) value of only one CKM element enters the      
theoretical predictions for weak tree-level decays, no GIM mechanism     
or unitarity condition has to be assumed. By matching theory and      
experiment the matrix element can be extracted {\it independently} of the number of  
generations \footnote{There is, however, one loop hole:   
In \cite{Lacker:2010zz} the possibility of lepton mixing reducing the accuracy of the  
determination of e.g. $V_{ud}$ was discussed. Since we use the more conservative error  
estimate from the PDG,  
our relative error is similar to the final error of Lacker and Menzel, who started 
with a more ambitious error for $V_{ud}$ in their analysis \cite{Lacker:2010zz}.}.   
Therefore, all tree-level constraints have      
the same impact on the $4\times 4$ matrix as they have on the       
$3\times 3$ one.     
      
We take the PDG values \cite{Amsler:2008zzb} for our analysis:     
 \begin{center}     
  \begin{tabular}{|c||c|c|c|}     
\hline     
   & absolute value  & relative error & direct measurement from \\ \hline     
$V_{ud}$ & $0.97418 \pm 0.00027$ & $0.028\%$& nuclear beta decay \\\hline     
$V_{us}$ & $0.2255  \pm 0.0019$  & $0.84\%$ & semi-leptonic K-decay\\\hline     
$V_{ub}$ & $0.00393 \pm 0.00036$ & $9.2\%$  & semi-leptonic B-decay\\\hline     
$V_{cd}$ & $0.230   \pm 0.011$   & $4.8\%$  & semi-leptonic D-decay\\\hline     
$V_{cs}$ & $1.04    \pm 0.06$    & $5.8\%$  & (semi-)leptonic D-decay\\\hline     
$V_{cb}$ & $0.0412  \pm 0.0011$  & $2.7\%$  & semi-leptonic B-decay\\\hline     
$V_{tb}$ & $>0.74$               &          & (single) top-production\\\hline     
  \end{tabular}     
 \end{center}     
In the following, we denote the absolute values in the table above as $|V_i|\pm\Delta V_i$.     
In addition to the above tree-level constraints there exists a direct bound on the CKM angle   
$$\gamma={\rm arg}\left( - \frac{V_{ud} V_{ub}^*}{V_{cd} V_{cb}^*}\right)\;.$$  
It can be extracted via the decays $B \to D K,D \pi$  
\cite{Gronau:1990ra,Gronau:1991dp,Atwood:1996ci,Giri:2003ty}.  
In principle the extraction of $\gamma$ might be affected by the   
presence of a fourth generation of fermions \cite{Kurimoto:1997ex}, but it was shown in \cite{Lacker2010} that these effects are negligible.   
Therefore, $\gamma$ gives direct information on the   
phases of the CKM matrix; with three CP violating phases present, this can provide a useful piece of information.  
We use the CKMfitter value from  \cite{CKMfitterhomepage} (update of \cite{Hocker:2001xe,Charles:2004jd})  
\begin{equation}  
 \gamma = 73^\circ { }_{+22^\circ}^{-25^\circ} \pm 2 ^\circ \, , 
\end{equation}  
where the last error accounts for the tiny additional uncertainty due to the   
additional fermion generation.

\subsubsection{Direct mass limits for the fourth family}   
   
The PDG \cite{Amsler:2008zzb} gives from direct searches the following mass limits for a fourth family   
\begin{eqnarray}      
m_{\nu_4}  & > & 80.5 ... 101.5   \, \mbox{GeV}  \, , 
\\   
m_{l_4}    & > & 100.8   \, \mbox{GeV}   \, , 
\\   
m_{b'}     & > & 128 ... 268   \, \mbox{GeV}  \, ,  
\\   
m_{t'}     & > & 256   \, \mbox{GeV}   \, . 
\end{eqnarray}      
The mass bound on the heavy neutrino depends on the type of neutrino (Dirac or Majorana) and whether   
one considers a coupling of the heavy neutrino to $e^-$, $\mu^-$ or $\tau^-$.   
It is interesting to note that {\it LEP results in combination with \cite{Reusser:1991ri} exclude a fourth   
stable neutrino with m $<$ 2400 GeV }  \cite{Amsler:2008zzb}.   
The quark mass bounds are obtained from direct searches at TeVatron \cite{Aaltonen:2007je,:2008nf}, which were     
recently updated \cite{Aaltonen:2009nr, CDFnote9446}   
\begin{equation}      
m_{b'}  >  338   \, \mbox{GeV}, \, \, \, \, \, \, \,       
m_{t'}  >  335   \, \mbox{GeV}.        
\end{equation}      
In \cite{Hung:2007ak} it was pointed out that in deriving these bounds assumptions about   
the couplings of the fourth generation have been made (in \cite{Aaltonen:2009nr} it is e.g. explicitly assumed that   
the $b'$ is shortlived and that it decays exclusively to $tW^-$, which corresponds to demanding   
$V_{ub'} \approx 0 \approx V_{cb'}$, $m_{b'} < m_{t'}$ and $V_{tb'}$ is not extremely small).  
Without these assumptions the mass bounds can be weaker, as the extraction of the masses has to be   
combined with the extraction of the CKM couplings. The inclusion of this dependence is beyond the scope of the current work.   
For some recent papers concerning the mass exctraction of leptons and quarks, see \cite{Carpenter:2010dt,Flacco:2010rg}. 
\\ 
In this work we investigate heavy quark masses in the range of 280 GeV to 650 GeV   
heavy charged lepton masses in the range of 100 GeV to 650 GeV   
and heavy neutrino masses in the range of 90 GeV to 650 GeV.    
Note that the triviality bound from unitarity of the $t't'$ $S$-wave  
scattering \cite{Chanowitz:1978mv} indicates a maximal $t'$ mass of around  
$504\;GeV$ \cite{Marciano:1989ns}. However, this estimate is based on tree-level  
expressions and while it seems prudent to treat too high quark masses with a grain  
of salt, one should not disregard higher masses based on this 
estimate alone. In this context it would be desireable to have e.g. a lattice  
study of the effect of  very heavy (fourth generation) quarks. 
%
%
%
\subsubsection{Electroweak constraints}   
\label{ewsect}  
We present here the expressions for the oblique electroweak   
$S,T$ and $U$ parameters \cite{Peskin:1991sw} in the presence   
of a fourth generation.   
They were originally defined as   
\begin{eqnarray}   
\alpha S & = &  4 e^2 \left. \frac{d}{dq^2}  \left[ \Pi_{33}(q^2) - \Pi_{3Q}(q^2) \right] \right|_{q^2 = 0}  \, , 
\label{define-S}  
\\   
 \alpha T & = &  \frac{e^2}{x_W \bar{x}_W M_Z^2}    \left[ \Pi_{11}(0) - \Pi_{33}(0) \right]  \, ,  
\label{define-T}  
\\   
\alpha U & = & 4 e^2 \left. \frac{d}{dq^2}  \left[ \Pi_{11}(q^2) - \Pi_{33}(q^2) \right] \right|_{q^2 = 0}  \, , 
\label{define-U}  
\end{eqnarray}   
with the electric coupling $\alpha$ and $e$, $\Pi_{xy}$ denotes the virtual self-energy contributions  
 to the weak gauge bosons and with the  Weinberg angle expressed as $x_W = \sin^2 \theta_W$ and $\bar{x}_W = 1 - x_W$.    
In the first paper of Ref. \cite{Peskin:1991sw}  $\bar{x}_W M_Z^2$  was approximated by $M_W^2$. 
The $T$ parameter is related to the famous $\rho$-parameter \cite{Veltman:1977kh,Chanowitz:1978uj,Chanowitz:1978mv}   
\begin{eqnarray}   
\rho & := & \frac{M_W^2}{\bar{x}_W M_Z^2} =: 1 + \Delta \rho \\   
     &  = & 1 + \alpha T  \, . 
\end{eqnarray}   
In practice, it turns out to be considerably simpler to reexpress the derivatives   
in $S$ and $U$ as differences   
\begin{equation}   
\left. \frac{d}{dq^2} \Pi_{XY}(q^2)  \right|_{q^2 = 0}   
\approx \frac{ \Pi_{XY}(M_Z^2) -  \Pi_{XY}(0)}{M_Z^2}   \, . 
\end{equation}   
This approximation works very well for $m_{new} \gg M_Z$ and it is used by the PDG \cite{Amsler:2008zzb}.   
We will use however the original definitions given in Eqs. (\ref{define-S}), (\ref{define-T}), (\ref{define-U})  
with $M_W^2 = \bar{x}_w M_Z^2$, because there are no 
correction terms and our expressions are exact.

\begin{figure}   
\begin{center}    
\includegraphics[width=0.9\linewidth, angle = 0]{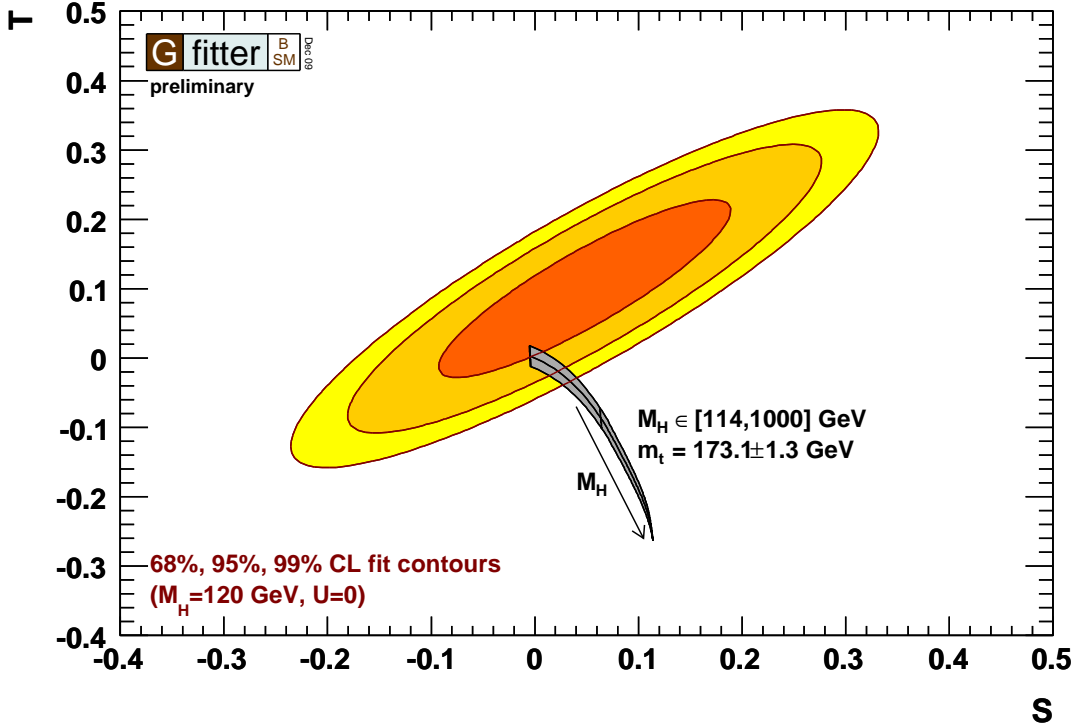}   
\end{center}  
\caption{Fit of the electroweak oblique parameters $S$ and $T$. The plot is taken from \cite{Gfitterhomepage}. \label{Gfit}}    
\end{figure}

\noindent   
Next only the new physics contributions to the $S$, $T$ and $U$ parameters will be considered,   
as the SM values of the oblique parameters are by definition set to zero. Fit results   
for the allowed regions of the $S$, $T$ and $U$ parameters are obtained e.g. by   
the PDG \cite{Amsler:2008zzb}, EWWG \cite{Alcaraz:2009jr}, Gfitter \cite{Flacher:2008zq} and  
most recent in \cite{Erler:2010sk}.  
Note that the more recent analyses \cite{Flacher:2008zq,Erler:2010sk} differ significantly from the old (November 2007) PDG version.  
Due to more refined experimental results and an improved theoretical understanding  
the best fit values shifted significantly towards higher values of $S$ and $T$, see Fig.~\ref{Gfit}   
for the Gfitter S-T ellipse \cite{Gfitterhomepage}; this somewhat relaxes the previously observed tension with an additional fermion generation.  
  
%
%
%
  
\noindent  
In the presence of a fourth generation the fermionic contribution to these parameters   
(before the necessary subtraction of the SM contribution) reads   
\begin{eqnarray}   
S & = &   \frac{N_c}{ 6 \pi} \sum \limits_{f=1}^4 \left[ 1 - \frac{1}{3} \ln\frac{m_{u_f}^2}{m_{d_f}^2} \right]   
       +  \frac{1  }{ 6 \pi} \sum \limits_{f=1}^4 \left[ 1 +    \ln\frac{m_{\nu_f}^2}{m_{l_f}^2} \right]  \, , 
\label{full-S}  
\\   
T & = & \frac{N_c}{16 \pi x_W \bar{x}_W M_Z^2}\left[  \sum \limits_{q=u,d,s,...,t',b'} m_q^2 -     
                \sum \limits_{f=1}^4 \sum \limits_{f'=1}^4  |V_{u_f d_{f'}}|^2    
                F_T( m_{u_f}^2, m_{d_{f'}}^2) \right]    
\\   
& + & \frac{1}{16 \pi x_W \bar{x}_W M_Z^2}\left[  \sum \limits_{l=\nu_e, e^-,..., \nu_4, l_4^-} m_l^2 -      
                \sum \limits_{f=1}^4 \sum \limits_{f'=1}^4  |V_{\nu_f l_{f'}}|^2    
                F_T(m_{\nu_f}^2, m_{l_{f'}}^2)  
\right]   \, , 
\label{full-T}  
\\  
U & = & \frac{N_c}{3 \pi} \left[   
\sum \limits_{f=1}^4 \sum \limits_{f'=1}^4  |V_{u_f d_{f'}}|^2  F_U( m_{u_f}^2, m_{d_{f'}}^2) - \frac{5}{6} \sum \limits_{f=1}^4 1  
 \right]  
\nonumber   
\\  
& +& \frac{1}{3 \pi} \left[   
\sum \limits_{f=1}^4 \sum \limits_{f'=1}^4   |V_{\nu_f l_{f'}}|^2   F_U ( m_{\nu_f}^2, m_{l_{f'}}^2) - \frac{5}{6} \sum \limits_{f=1}^4 1  
 \right] \, . 
\label{full-U}  
\end{eqnarray}   
$u_f$ denotes the up-type quark of the $f$th generation,    
$d_f$ the down-type quark of the $f$th generation, $l_f$ the charged lepton of the $f$th generation and $\nu_f$ the neutrino of   
the $f$th generation.   
We have used the following functions  
\begin{eqnarray}  
F_T(m_1^2, m_2^2) & := & 2\frac{m_1^2 m_2^2}{m_1^2 - m_2^2} \ln \frac{m_1^2}{ m_2^2} \, , 
\\  
F_U(m_1^2, m_2^2) & := & 2 \frac{m_1^2 m_2^2}{(m_1^2 - m_2^2)^2}  
                         + \left(  
  \frac{m_1^2 + m_2^2}{2 (m_1^2 - m_2^2)}  
- \frac{m_1^2 m_2^2 (m_1^2 + m_2^2)}{(m_1^2 - m_2^2)^3}  
                           \right) \ln \frac{m_1^2}{ m_2^2} \, . 
\end{eqnarray}  
Both functions are symmetric in their arguments.  
\\  
The formula for $S$ is very well known - see e.g.   
\cite{Peskin:1991sw,He:2001tp,Kribs:2007nz,Chanowitz:2009mz}.  
Using instead the PDG definition we would obtain the following corrections terms to $S$ for heavy quark masses ($m_q^2 \gg M_Z^2$)   
\begin{eqnarray}   
S_q^{corr.} &= &  \frac{N_c}{ 6 \pi} \left[ \frac{M_Z^2}{3 m_{b'}^2} \left( - \frac12 + \frac23 x_W  - \frac49     x_W^2 \right)   
               + \frac{M_Z^2}{3 m_{t'}^2} \left( - \frac12 + \frac43 x_W - \frac{16}{9} x_W^2 \right) \right] \, , 
\\   
S_l^{corr.} &= & - \frac{1  }{ 6 \pi} \left[  \frac{M_Z^2}{6 m_{\nu_4}^2}    
                 + \frac{M_Z^2}{2 m_{l_4}^2} \left( - \frac13 + \frac13 x_W - 3 x_W^2 \right) \right]  \, , 
      \end{eqnarray}   
which are very small for the allowed mass ranges of the fourth family members.  
In the parameter $T$ no mixing was usually assumed. We give here the full CKM and PMNS  
dependence. Our expression for $T$ in Eq. (\ref{full-T}) agrees with the one quoted   
in \cite{Chanowitz:2009mz}, if we make the same assumptions (only $4-3$ mixing or   
$4-3$ and $4-2$ mixing is considered).   
\\   
By defining the SM3 values for $S$ and $T$ as zero, we only need to  
take the additional contributions due to the fourth   
generation into account. Keeping the full, previously   
neglected CKM dependences, we obtain   
\begin{eqnarray} \label{S4eq}  
S_4 & = &  \frac{1}{ 3 \pi} \left[ 2 + \ln\frac{m_{b'}\,  m_{\nu_4}}{m_{t'} \, m_{l_4}}  
\right]  \, , 
\\   
T_4 & = & \frac{N_c}{16 \pi x_w \bar{x}_w M_Z^2} \left[  m_{b'}^2 + m_{t'}^2 -      
                \sum \limits_{f=1}^4 \sum \limits_{f'=1}^4  |V_{u_f d_f'}|^2    
                  F_T \left( m_{u_f}^2, m_{d_{f'}}^2 \right)   
               +  F_T \left( m_{t}^2  , m_{b}^2      \right)  
\right]    
\nonumber \\ \label{T4eq}  
&& +  \frac{1}{16 \pi x_w \bar{x}_w M_Z^2} \left[  m_{l_4}^2 + m_{\nu_4}^2 -      
                \sum \limits_{f=1}^4 \sum \limits_{f'=1}^4  |V_{\nu_f l_f'}|^2    
                F_T \left( m_{\nu_f}^2, m_{l_{f'}}^2 \right)  
\right]   \, , 
\\   
U_4 & = &  - \frac{N_c}{6\pi }\left[    
 |V_{t'd}|^2 \ln \frac{m_{t'}^2}{m_d^2}   
+|V_{t's}|^2 \ln \frac{m_{t'}^2}{m_s^2}  
+|V_{t'b}|^2 \ln \frac{m_{t'}^2}{m_b^2}   
+|V_{ub'}|^2 \ln \frac{m_{b'}^2}{m_u^2}   
+|V_{cb'}|^2 \ln \frac{m_{b'}^2}{m_c^2}   
\right.  
\nonumber   
\\   
&& \hspace*{30pt}   
\left.  
- 2 |V_{tb' }|^2 F_U \left( m_t^2   , m_{b'}^2 \right)  
- 2 |V_{t'b'}|^2 F_U \left( m_{t'}^2, m_{b'}^2 \right) \right]  
\nonumber   
\\   
&&  
 - \frac{1}{6\pi }\left[    
  |V_{{\nu_4}e}|^2 \ln \frac{m_{{\nu_4}}^2}{m_e^2}   
+ |V_{{\nu_4}\mu}|^2 \ln \frac{m_{{\nu_4}}^2}{m_\mu^2}  
+ |V_{{\nu_4}\tau}|^2 \ln \frac{m_{{\nu_4}}^2}{m_\tau^2}   
+ |V_{\nu_e{l_4}}|^2 \ln \frac{m_{{l_4}}^2}{m_{\nu_e}^2}   
+ |V_{\nu_\mu{l_4}}|^2 \ln \frac{m_{{l_4}}^2}{m_{\nu_\mu}^2}   
\right.  
\nonumber   
\\   
&& \hspace*{30pt}   
\left.  
+   |V_{\mu_\tau{l_4} }|^2 \ln  \frac{ m_{{l_4}}^2}{m_{\mu_\tau}^2}  
- 2 |V_{{\nu_4}{l_4}}|^2 F_U \left( m_{{\nu_4}}^2, m_{{l_4}}^2 \right) \right]  
- \frac{10}{9 \pi} + U_{SM3}  \, . 
\nonumber   
\\   
\label{U4eq}  
\end{eqnarray}   
$S_4$ has a large positive contribution of about $0.21$ which is independent   
of the parameters (masses and mixing) of the model. This value can, however,   
be diminished by the second logarithmic term that depends on the fermion masses.   
\\   
In the SM3 the only significant contribution to $T$ reads  
\begin{eqnarray}  
T_4 & = & \frac{N_c}{16 \pi x_w \bar{x}_w M_Z^2} \left[ m_{t}^2 - F_T \left( m_{t}^2  , m_{b}^2      \right)  
\right]   \, . 
\end{eqnarray}  
Here safely $V_{tb} = 1$ can be assumed. In the SM4, however,  $V_{tb}$ can in 
principle differ significantly from one,   
therefore we have the correction term in ``$ + F_T \left( m_{t}^2  , m_{b}^2      \right) $''   
in the formula for $T_4$.  
We also have included all previously neglected mixing terms within the SM3 particles. 
In principle we also should correct for the charm-strange  
contribution and for the up-down contribution with 
``$ + F_T \left( m_{u}^2  , m_{d}^2      \right) + F_T \left( m_{c}^2  , m_{s}^2      \right) $'' ,  
but their numerical effect is considerably below one per mille of the top-bottom contribution,  
so we do not show these two additional correction terms in the formula for $T_4$.\\  
With the help of the $S$ and $T$ parameter Chanowitz \cite{Chanowitz:2009mz} could exclude  
the three parameter sets,  which we gave in \cite{Bobrowski:2009ng} as an example for a very  
large mixing between the third and the fourth generation; these sets have passed all bounds set by  
precision flavor observables.  
We confirm the numbers from table I in \cite{Chanowitz:2009mz}. We also tested the approximation of   
taking only $3-4$ mixing into account:  Comparing with the full CKM dependence the  
differences are below 6 $\%$ for these three parameter sets.   
\\  
To simplify the expression for $U$ we approximated  
\begin{equation}  
F_U (m_1^2, m_2^2) \approx - \frac{1}{2} \ln \frac{m_1^2}{m_2^2} \, \, \, \, \, \,  
\, \, \, \mbox{for} \, \, \, \, \, m_1^2 \ll m_2^2 \, . 
\end{equation}  
Moreover we have only shown the contributions of the 4th family explicitly in Eq. (\ref{U4eq}), the  
previously neglected rest is denoted by $U_{SM3}$.  
It will be interesting to   
see in a future analysis, whether the large logarithms in the lepton sector will lead to strong constraints on  
the PMNS-matrix.   
In the literature it is typically assumed that $U$ is very small, see \cite{He:2001tp} for a notable exception.   
To our knowledge we incorporate for the first time the full   
CKM dependence in $U$.    
We find that arbitrary values for mixing and mass parameters could in principle generate values as  
large as $7.5$ for $U$\footnote{We did not check whether this is the largest possible value.}.   
If one only takes into account mixing parameters that pass the   
tree-level flavor constraints still values of $ {\cal O} (0.1)$ seem to be possible;   
however, in this case we observe a simultaneous blow up of the $T$ parameter.   
For $T<0.4$, $U$ does not exceed $0.06$. Note that, while still small, this value is larger than the $0.02$   
effect expected without flavor mixing \cite{Kribs:2007nz}.  
  
At this point a few comments are appropriate: first, we would like to point out that our   
implementation of the $S$ and $T$ parameter is not ``exact'' from the SM4 point of view.  
In principle one would have to perform a full reanalysis of all electroweak data  
from the SM4 perspective to fit the new values of $S$ and $T$, as advocated for   
in \cite{Chanowitz:2009mz}.   
This is, of course, beyond the scope of the present work and it is generally  
accepted that a large deviation of the oblique parameters from their SM values cannot   
be accommodated in models that do not introduce new particles coupling to fermions   
\cite{Amsler:2008zzb}.  
Secondly, we will henceforth neglect the effect of lepton mixing due to   
a non-trivial modification of the PMNS matrix --- the off diagonal elements including the fourth   
neutrino are in any case required to be small, see \cite{Lacker:2010zz}.  
  
\vspace{0.3cm}   
\noindent   
As a first step in our analysis, we only take the tree-level constraints on $V_{CKM4}$   
into account and investigate the parameter ranges that pass the $S-T$ test at the $95\%$  
confidence level\footnote{For the final investigation of the allowed parameter range of a fourth fermion family 
we will use the $99\%$ confidence level.}.  
The following values for the SM fit of the oblique parameters are used   
\cite{Erler:2010sk}\footnote{Note   
that this fit is the most restrictive one currently available. Using instead the results from  
Gfitter a little more space is left for a fourth family. We simply decided to use the most recent numbers.}  
\begin{align}  
 S_{\rm best fit}&= 0.03 \, ,&&\sigma_S=0.09 \, , \\  
 T_{\rm best fit}&= 0.07 \, ,&&\sigma_T=0.08 \, , \\  
\rho_{corr}&=0.867\;,  
\end{align}  
where $\sigma_x$ gives the standard deviation of $x$.  
The $S$ and $T$ parameters are not independent quantities;  
the strength of this correlation is given by $\rho_{corr}$. 
 
As our work focuses on the flavor aspects of the 4th generation  
scenario, a short comment on the famous S-T-ellipses seems to be in order.  
If the $U$ parameter stays close to zero for some physics model 
it is feasible to set $U=0$ and work with $S$ and $T$ alone.  
In this case the probability distribution of $S$ and $T$ reduces  
to a two-dimensional Gaussian distribution  
in the S-T plane centered on the best fit values. Due to the  
different $\sigma_x$ and due to the strong correlation the  
distribution is essentially squeezed and rotated. Hence, the  
``equiprobability'' lines are no longer circles but ellipses. 
In fact the two dimensional case is somewhat special as  
the problem of finding the ellipse encircling an area corresponding to 
certain probability $P$ can be solved analytically. The equation determining 
the contour for a given confidence level $CL$ is then given by  
\begin{align} 
 \begin{pmatrix} 
  S-S_{\rm best\;fit}\\ 
  T-T_{\rm best\;fit} 
 \end{pmatrix}^T 
  \begin{pmatrix} 
    \sigma_S \sigma_S&\sigma_S \sigma_T \rho \\ 
    \sigma_S \sigma_T \rho&    \sigma_T \sigma_T 
 \end{pmatrix}^{-1} 
  \begin{pmatrix} 
  S-S_{\rm best\;fit}\\ 
  T-T_{\rm best\;fit} 
 \end{pmatrix} = -2 \ln(1-CL)\;. 
\end{align} 
\label{STUFit}

Already at that stage we find some interesting results:   
\begin{enumerate}   
\item $S-T$ test with `no leptons' + `no $V_{CKM}$' 
      \\   
      We do not take into account leptons as well as mixing of the quarks.  
      By neglecting the leptonic contribution one can, of course, not make any conclusions   
      as to how restrictive the oblique parameters are. However, we still find it   
      instructive to consider the effect of the various contributions in the S-T plane  
      individually.   
      The scatter plot shows the accessable region within the $95\%$ CL ellipse of \cite{Erler:2010sk}.  
      We find, as expected that the masses of the fourth quark generation can not be   
      degenerate if they fulfill the constraints from the oblique parameters. The   
      necessary mass difference is of the order of 50 GeV as stated in \cite{Kribs:2007nz}.     
\begin{center}    
\includegraphics[width=0.6\linewidth, angle = 0]{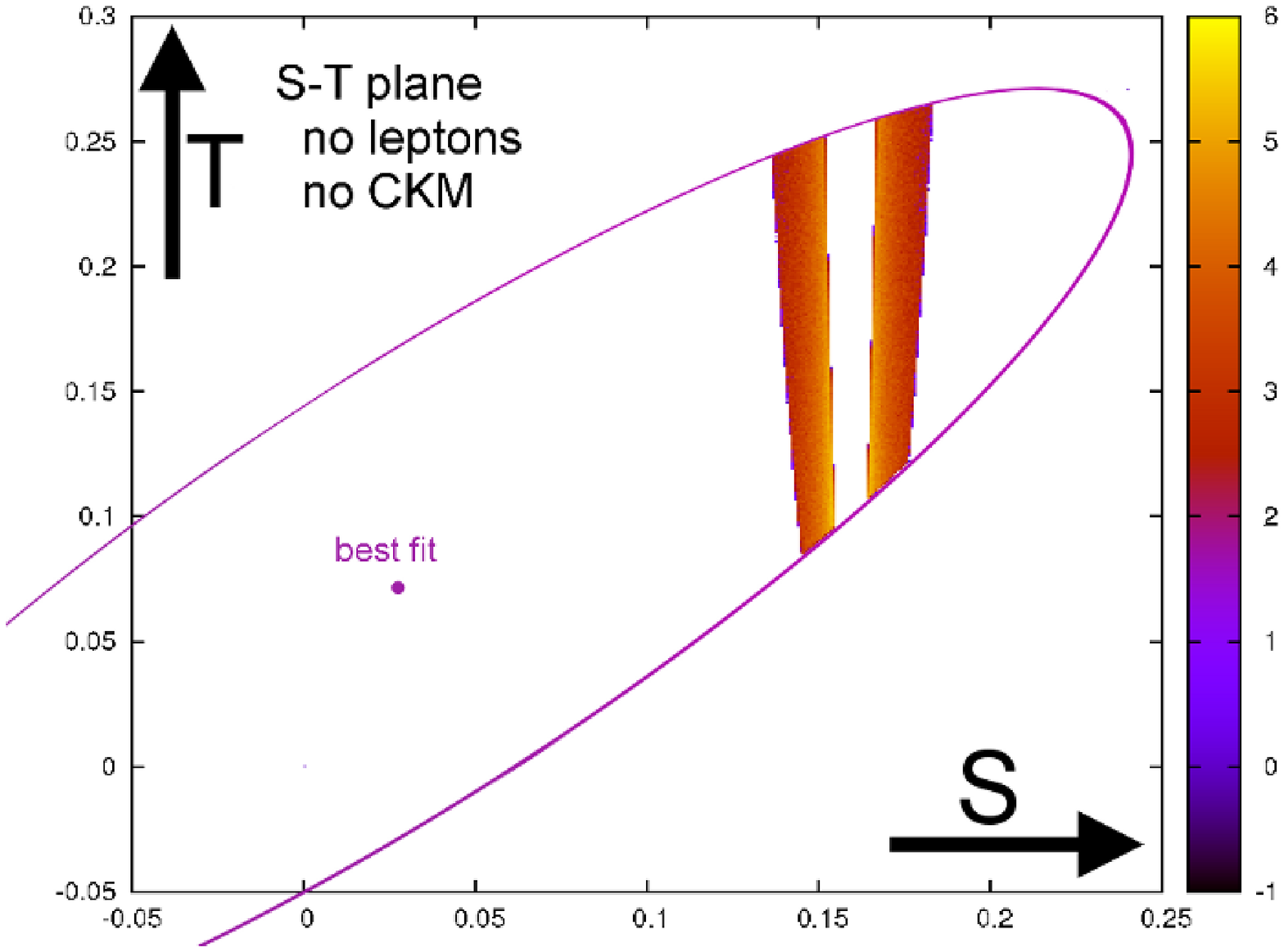}   
\end{center}    
%
   
\item $S-T$ test with `with leptons'  + `no $V_{CKM}$'   
      \\   
      Next, we include the leptonic contributions (without lepton mixing) and still neglect CKM   
      mixing. In this case also degenerate values of the quark masses of the fourth generation   
      are in principle not excluded;  
however, this would require a significant mass gap in the lepton 
doublet to increase $T$ (and preferably reduce $S$). 
 
\begin{center}    
\includegraphics[width=0.6\linewidth, angle = 0]{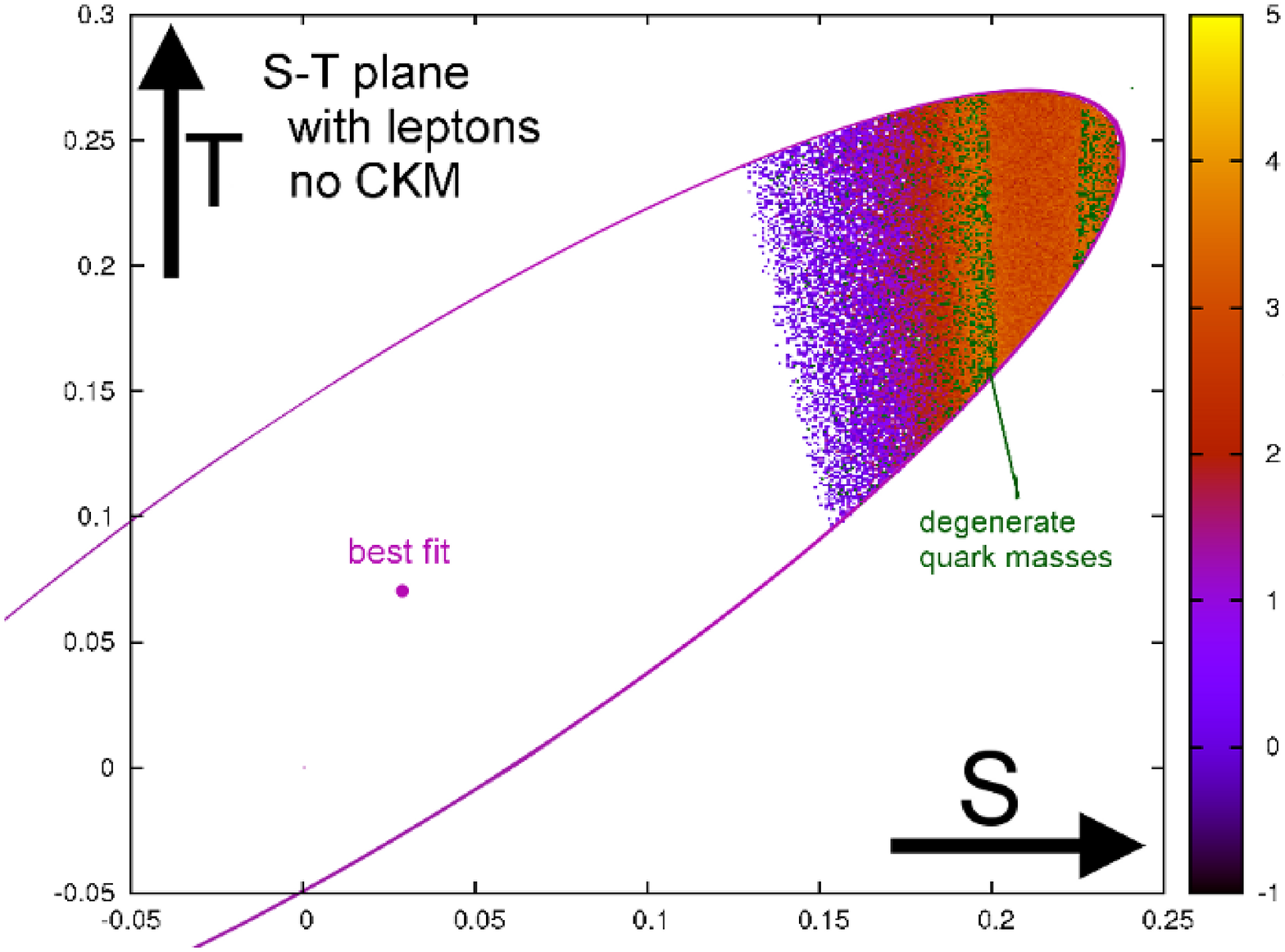}   
\end{center}    
%
The dark green points indicate values not in conflict with a degeneracy of the quark masses of the fourth generation.   
\item $S-T$ test with `no leptons'  + `with $V_{CKM}$'  
\\         
To study the ``unperturbed'' effect of a non trivial CKM structure of the fourth generation, let us    
discard again the leptons for the moment. The modified CKM structure results in an increase of the   
$T$ parameter without changing $S$, cf.~Eqs.~\eqref{S4eq} and \eqref{T4eq}.   
\begin{center}    
\includegraphics[width=0.6\linewidth, angle =0]{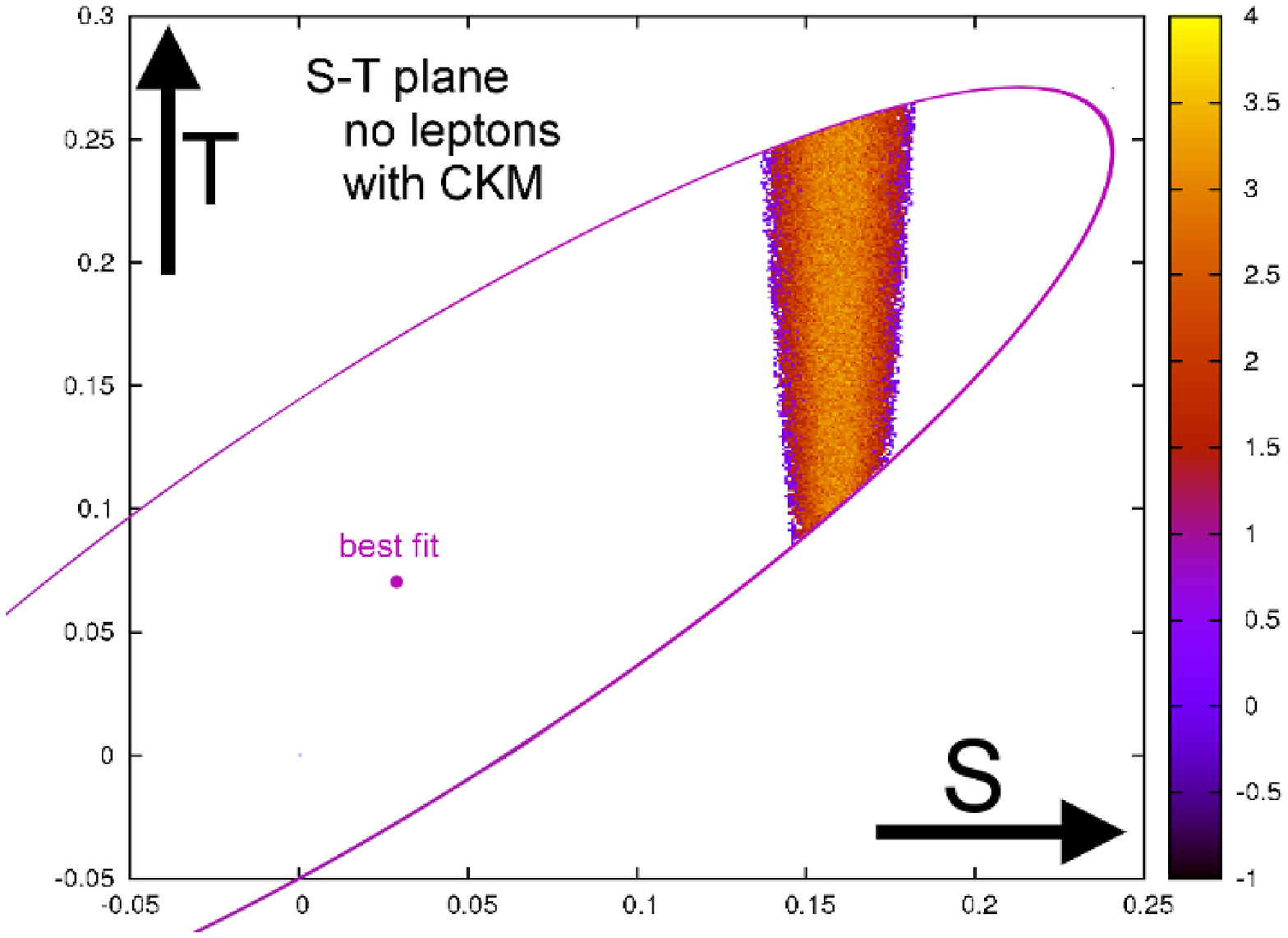}      
\end{center}    
%
In this scenario an increase of $T$ can originate either from a quark mass splitting or from nonzero mixing of the fourth generation with the SM quarks.
However, a mass splitting must also induce a tiny (logarithmical) contribution
to $S$; so the central area which could not be reached in scenario '1'
corresponds to nonzero mixing and tiny mass splittings.

\item $S-T$ test with `with leptons'  + `with $V_{CKM}$' \\   
Here, we use the full expressions for $S$ and $T$ including both, leptons (without mixing) and CKM mixing.   
In this scenario we find a maximally allowed mass splitting of   
$|m_{t'}-m_{b'}|<80$  GeV for quarks and $|m_{l_4}-m_{\nu_4}|<140 $ GeV for   
leptons.   
\\  
Note that this splitting was also observed by the Gfitter group \cite{Gfitterhomepage};  
their fits also show a minimal required mass splitting as they do not take the  
possible effects of a nontrivial CKM structure into account. 
Due to the effects of quark mixing, we do not find a  lower bound for the splitting. 
In fact, a simultaneous degeneracy of quark and lepton masses is {\it not} excluded,   
even though the $S$ parameter favors larger $t'$ and $l_4$ masses.   
\begin{center}    
\includegraphics[width=0.6\linewidth, angle = 0]{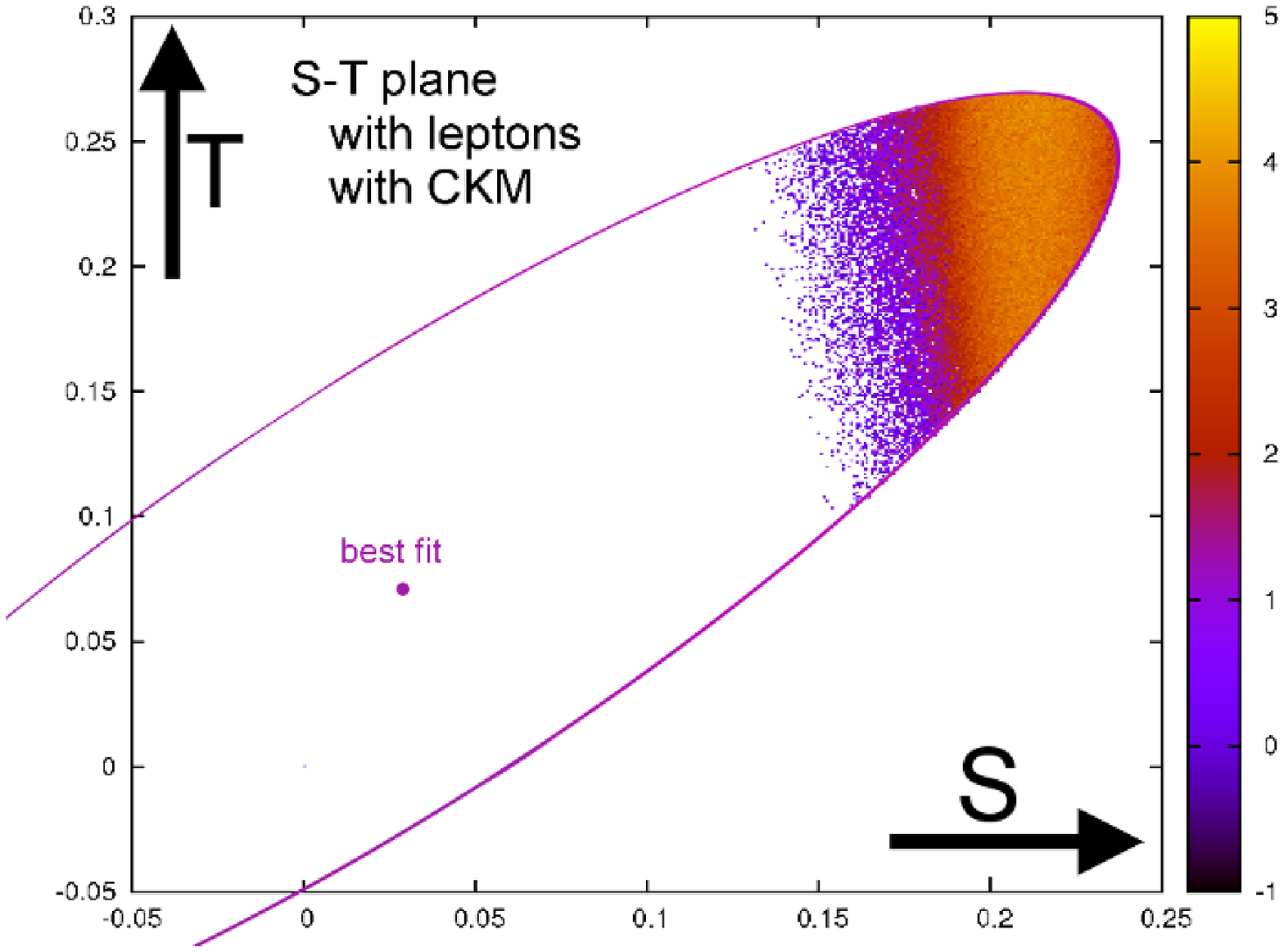}   
\end{center}    
%
   
\item For completeness we also show the $S$-$U$ plane using the exact expression for $U$ and $S$.  
\begin{center}    
\includegraphics[width=0.6\linewidth, angle = 0]{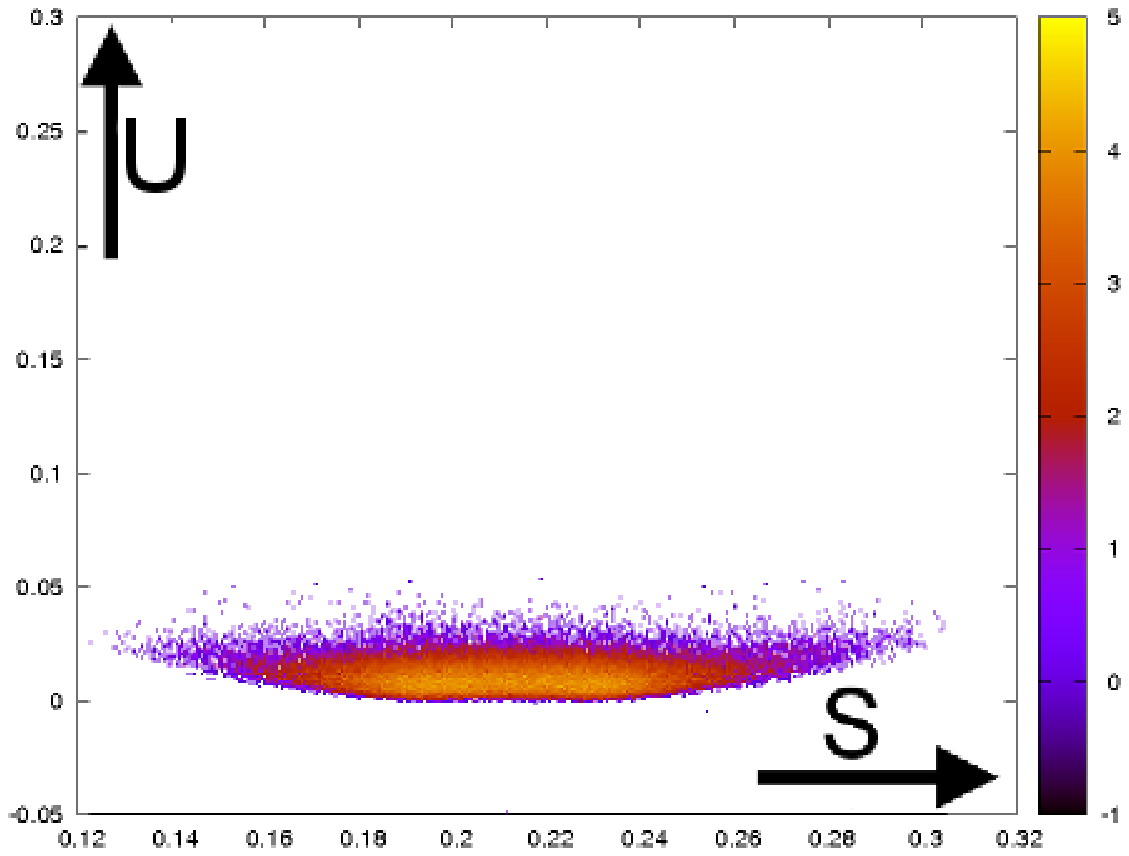}   
\end{center}    
%
\end{enumerate}   
   
Summarizing our investigation of the $S$, $T$ and $U$ parameters we get the following results:   
\begin{itemize}   
\item U is not a priori small; only after the constraints on the quark mixing and the $T$ parameter are used the   
maximal value for $U$ is reduced below $0.06$.  
\item The quarks of a 4th generation can be degenerated without violating   
      the $95\%$ CL constraints from electroweak precision  
      observables\footnote{For the $99\%$ CL this statement will even hold stronger.}.   
      However, this requires taking into account the   
      effects of the non trivial flavor sector on $T$, i.e. mixing of the $4$th generation   
      fermions, or a sufficiently   
      large mass splitting in the lepton sector. Note that, at first glance, this result   
      seems to be in direct conflict with the standard statement that a degenerate fourth   
      generation is excluded at the 6 $\sigma$ level \cite{Amsler:2008zzb} by virtue of the   
      $S$ parameter. However, this statement always tacitly assumed a trivial CKM structure.  
      The CKM factors in Eq.~\eqref{T4eq} can lead to $T>0$ even if both lepton and quark masses  
      are degenerate. However,  we did not investigate the effect of the   
      $Z\to \bar b b$ vertex, which tends to favor small or no mixing. Still one can conclude   
      that the situation for tiny mass splittings or even degenerate masses  
      drastically improves once mixing is taken into account.

%
\end{itemize}   
   
\noindent   
Finally, we also have to (re)consider the contribution of the Higgs particle, since in   
the presence of a fourth family higher values of the Higgs mass may be possible \cite{Peskin:1991sw}. The   
correction terms to the $S$, $T$ and $U$ parameters read   
      \begin{eqnarray}   
      S_H & = &   \frac{1}{12 \pi} \ln \frac{M_H^2}{(117 \, \mbox{GeV})^2}   \, , 
      \\   
      T_H & = & - \frac{3}{16 \pi \bar{x}_W} \ln \frac{M_H^2}{(117 \, \mbox{GeV})^2} \, ,   
      \\   
      U_H & \approx & 0  \, . 
      \end{eqnarray}   
Using that form for the Higgs contributions we implicitly subtract the used value for the Higgs mass in the   
fit ($117\;\rm GeV$) and add ``our'' value.  
 
A heavier Higgs increases $S$ and lowers $T$. Instead of adding $S_H$ and $T_H$ to our  
values for $S$ and $T$, we subtracted the Higgs contributions from the fit values to  
make the diagram easier to understand. So we shift the ellipse and not our data sets.  
We investigate the $S$ and $T$ parameters for three values of  
the Higgs mass: 117 GeV, 250 GeV and 500 GeV.

\begin{center}    
\includegraphics[width=0.6\linewidth, angle = 0]{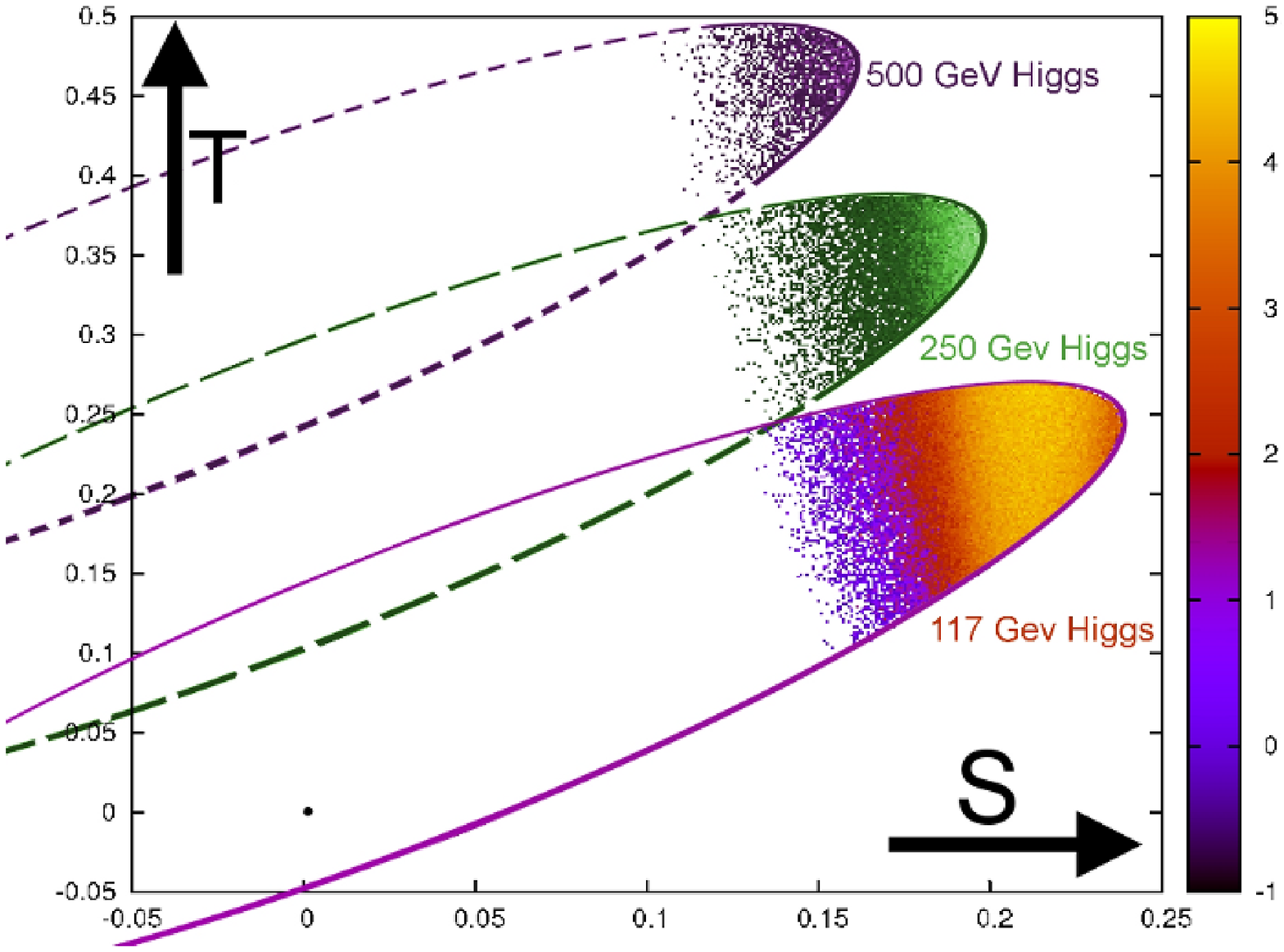}   
\end{center}    
The decrease in $T$ is welcome, as it allows  even bigger mass splitting (or alternatively larger mixing);   
however, the simultaneous increase of $S$ due to the heavy Higgs  
completely seems to neutralize or even reverse this effect. Hence, as stated  
recently in \cite{Erler:2010sk}   
very large values for the mass of a SM-like Higgs are clearly not favored.  
However, for a 250 GeV Higgs scenario  
the origin (SM3) is outside the ellipse, whereas some SM4 points are inside and thus more likely.

%
%
%
%
%
%
%
  
\subsection{Flavor physics  constraints --- FCNC processes}  
  
  After addressing the electroweak bounds, we turn to the constraints   
  imposed   
  by precision observables of flavor physics involving a FCNC.  
  One can hope to impose severe constraints on the model by utilizing   
  information   
  from such processes as it is well known that the weak interaction bypasses   
  the Appelquist-Carazzone decoupling theorem \cite{Appelquist:1974tg}; thus, FCNC processes   
  are very sensitive to contributions of new physics.  
  
  However, the selection of flavor physics bounds on a hypothetical fourth   
  family is a non trivial issue. The reason for this is the fact that some   
  processes known for being theoretically or experimentally very clean,   
  may in fact specifically require the SM3 setup. Hence, it is always   
  necessary to check, if a specific feature (of the SM3) crucial   
  e.g.~ for the data analysis is preserved in the fourth generation   
  extension. If this is not the case, it may either be necessary   
  to repeat the analysis without some SM3 simplification ---   
  much like the need to give up $3\times 3$ unitarity ---  
  or the whole process may not even be feasible anymore.  
  
\begin{figure}   
\begin{minipage}{0.475\textwidth}  
 \begin{center}  
 \includegraphics[width=1.0\textwidth]{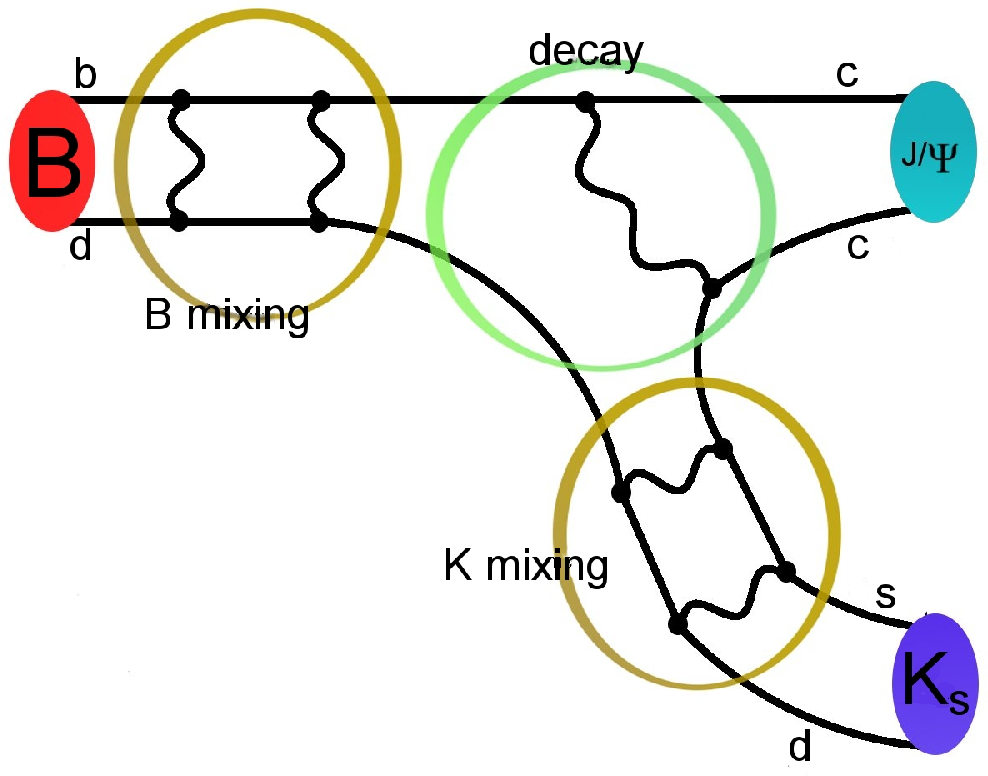}  
\end{center}  
\end{minipage}  
\begin{minipage}{0.475\textwidth}  
\begin{center}  
 \includegraphics[width=1.0\textwidth]{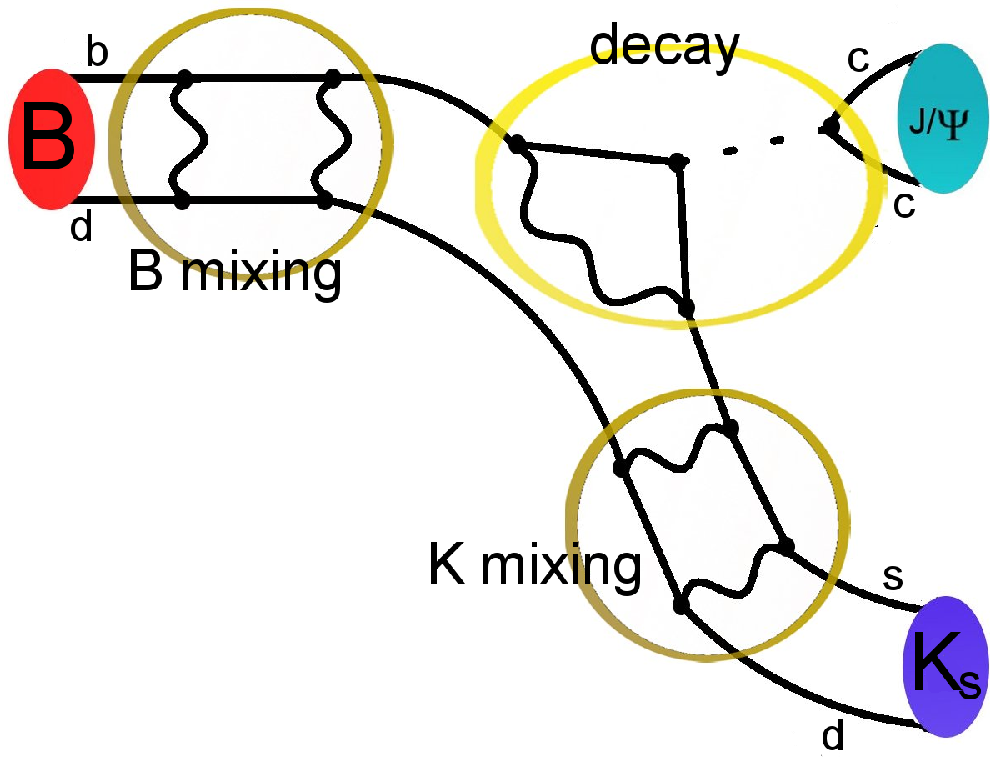}  
\end{center}  
\end{minipage}  
\caption{Schematic diagrams for the necessary ingredients of the  
         golden plated channel $B_d \to J/\Psi K_s $: B-mixing, Kaon mixing   
         and the decay process itself.  
         The left panel shows the tree-level and the right panel the   
         penguin mediated decay.  
         The dashed line represents any current capable of creating a   
         $J/\Psi$, e.g.~two gluons.   
         \label{GoldenMode} }  
\end{figure}  
  
  As an example for how unexpected complications may arise   
  (see also \cite{Chiang:2009ev} for a more detailed discussion),  
  we discuss the so-called golden plated channel for the  
  determination of the Standard Model CKM   
  angle $\beta$: $B_d \to J/\Psi K_s$ \cite{Bigi:1981qs}. This channel is   
  renowned for being theoretically very clean (in the SM3).  
  Since the decay process is tree level dominated,  
  it is usually taken for granted  
  that the contribution of the fourth generation quarks to the decay   
  is generally small. Therefore the SM4 could, in principle, be an explanation for  
  discrepancies of the measurement of $\sin(2\beta)$ in the     
  $B_d \to J/\Psi K_s$ and $B_d \to \Phi K_s$, as $B_d \to \Phi K_s$  
  is penguin dominated and as such more sensitive to new physics effects,  
  see \cite{Soni:2009zz}  for a more detailed version of this argument.  
  However, it turns out that this elegant picture  
  of the consequences of the fourth generation is, unfortunately, too  
  simple. The reason for this is the following:  
  $\sin(2\beta)$ is extracted via time-dependent CP asymmetries.  
  The necessary ingredients are (i) $B_d$ mixing, (ii) Kaon mixing and   
  the (iii) decay process itself, see Fig.~\ref{GoldenMode} for a schematic picture  
  of the relevant subprocesses.  
  There are in fact two decay processes, the tree-level decay and the   
  top mediated penguin decay ($c$ and $u$ penguin are expected to be   
  tiny). However, the beauty of this process in the SM3 is that  
  the tree level decay and the $t$ penguin have (to a fantastic accuracy)  
  the same CKM phase. Hence, it is not necessary to take into account   
  e.g.~different hadronization effects as they will only modify the overall  
  amplitude but not the phase. Adding an additional generation has two  
  new, separate effects. First of all, the expressions for the box diagram  
  changes (see formulae below), so that additional CKM factors contribute; therefore,  
  instead of the CKM angle $\beta$ a different combination of CKM angles 
  can be extracted from this process. This is, however,   
  not a problem as one could still use $B_d \to J/\Psi K_s$ to constrain   
  the SM4. The real problem is the simultaneous modification of the   
  penguin diagram by the $t'$ loop. As the $t'$ will introduce some  
  new virtually unconstrained phase, penguin and tree decay now have  
  {\it different} CKM phases; the fourth generation   
  introduces a mismatch between tree and penguin decay, which   
  makes taking into account hadronization and QCD corrections   
  mandatory. Therefore, it is not clear what quantity can be   
  extracted from time-dependent CP asymmetries in $B_d \to J/\Psi K_s$ in the  
  SM4 scenario. This, of course, limits the usefulness  
  of this process for constraining the parameters of the model.  
    
  As the above example shows, not all processes can be used to  
  obtain limits on the parameters of the SM4 and one has to be careful not   
  to make use of a bound whose experimental input essentially requires   
  the SM3 setup or some SM3 specific feature.   

\subsubsection{FCNC constraints with no sensitivity to lepton mixing}   
   
  Another issue is the sensitivity of some processes to the properties   
  of the leptons. Since we do not take mixing in the lepton sector into   
  account and in fact assume lepton number conservation, quantities  
  that are insensitive to the precise structure of the lepton sector  
  are of course advantageous.


\paragraph{We first consider the mixing of the $K$-, the $D$-, the $B_d$- and the $B_s$-system.}   
For completeness we repeat the relevant formulae already given in \cite{Bobrowski:2009ng}.   
The virtual part of the box diagram (here e.g. for $B_d$-mixing)   
      \begin{center}     
      \includegraphics[width=0.9\linewidth]{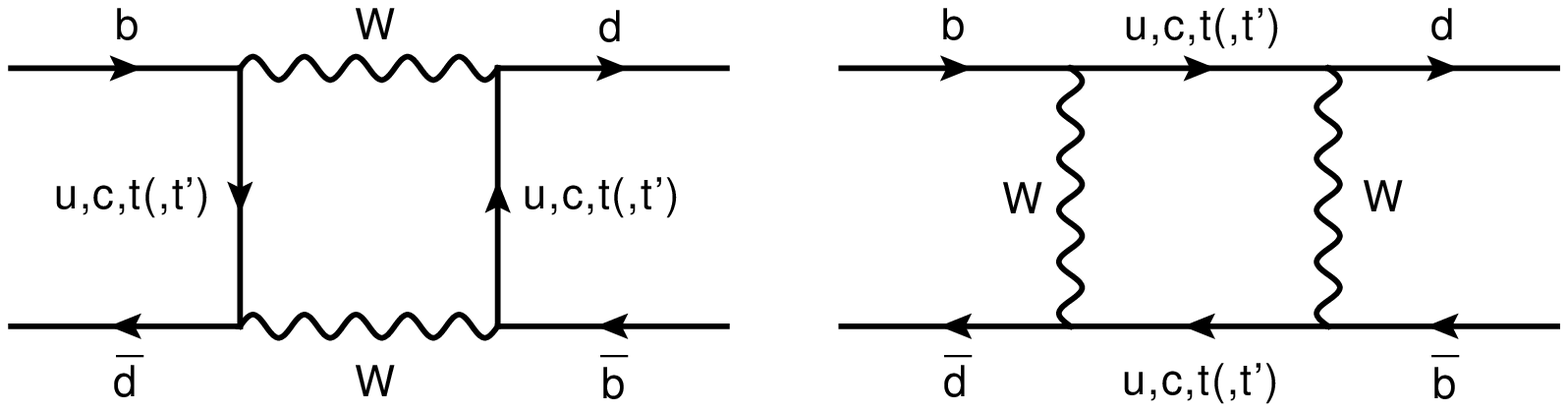}      
      \end{center}     
is encoded in $M_{12}$, which is very sensitive to new physics contributions.   
It is related to the mass difference of the heavy and light neutral mass eigenstate via      
      \begin{equation}      
      \Delta M = M_{B_H} - M_{B_L} = 2 | M_{12}|\, .      
      \end{equation}      
In the SM3 one obtains the following relations     
      \begin{eqnarray}      
      M_{12}^{K^0} & \propto & \eta_{cc} \left(\lambda_c^{K^0}\right)^2 S_0(x_c)       
                         + 2 \eta_{ct} \lambda_c^{K^0} \lambda_t^{K^0} S(x_c,x_t)       
                         + \eta_{tt} \left(\lambda_t^{K^0}\right)^2 S_0(x_t) \,  ,     
      \\      
      M_{12}^{B_d} & \propto & \eta_{tt} \left(\lambda_t^{B_d}\right)^2 S_0(x_t) \, ,     
      \\      
      M_{12}^{B_s} & \propto & \eta_{tt} \left(\lambda_t^{B_s}\right)^2 S_0(x_t) \, ,     
        \end{eqnarray}      
      with the Inami-Lim functions \cite{Inami:1980fz}     
     \begin{eqnarray}      
     S_0(x) & = & \frac{4x - 11 x^2 + x^3}{4(1-x)^2} - \frac{3 x^3 \ln[x]}{2 (1-x)^3} \, ,     
     \\      
     S(x,y) & = & x y  \left[      
                  \frac{1}{y-x} \left( \frac14 + \frac32 \frac{1}{1-y} - \frac34 \frac{1}{(1-y)^2} \right) \ln[y]      
        \right. \nonumber \\       
          &&  \left.   + \frac{1}{x-y} \left( \frac14 + \frac32 \frac{1}{1-x} - \frac34 \frac{1}{(1-x)^2} \right) \ln[x]      
              - \frac34  \frac{1}{1-x} \frac{1}{1-y} \right] \, ,     
     \label{box2} 
     \end{eqnarray}      
     where $ x_{c,t}  =  \frac{m_{c,t}^2}{M_W^2}$,  the CKM elements      
     \begin{equation}      
     \lambda_{x}^{K^0} = V_{xd} V_{xs}^*,  \, \, \, \, \, \, \,        
     \lambda_{x}^{B_d} = V_{xd} V_{xb}^*,  \, \, \, \, \, \, \,        
     \lambda_{x}^{B_s} = V_{xs} V_{xb}^*          
     \end{equation}      
     and the QCD corrections \cite{Buras:1990fn,Herrlich:1993yv,Herrlich:1996vf}     
     \begin{equation}      
     \eta_{cc}= 1.38   \pm 0.3    ,  \, \, \, \, \, \, \,       
     \eta_{ct}= 0.47   \pm 0.04   ,  \, \, \, \, \, \, \,       
     \eta_{tt}= 0.5765 \pm 0.0065 .      
     \end{equation}      
     The full expressions for $M_{12}$ can be found e.g. in \cite{Buras:1990fn,Lenz:2006hd}.      
     In deriving these expressions the unitarity of the $3\times 3$ matrix   
     was explicitly used, i.e.     
     \begin{equation}      
     \lambda_u^X + \lambda_c^X + \lambda_t^X = 0 \, .      
     \end{equation}      
     Moreover, in the $B$-system the CKM elements of the different internal   
     quark       
     contributions are all roughly of the same size. Only the top   
     contribution, which has by far the largest      
     value of the Inami-Lim functions, survives. This is not the case in the   
     $K$-system. Here      
     the top contribution is CKM suppressed, while the kinematically   
     suppressed charm terms are      
     CKM favored. Therefore, both have to be taken into account.      
     More information about the mixing of neutral mesons can be found e.g. in   
     \cite{Lenz:2006hd, Anikeev:2001rk}.     
     \\      
     We define the parameter $\Delta$       
     as the ratio of the new physics model prediction   
     (in our case SM4) for a generic observable to the SM3 theory value;   
     thus, it quantifies the  deviation from the standard model \cite{Lenz:2006hd}.  
     For $M_{12}$ one would then define:       
      \begin{equation}      
      \Delta := \frac{M_{12}^{SM4}}{M_{12}^{SM3}} = |\Delta|   
                e^{i\phi^{\Delta}}\, .      
      \end{equation}      
    This representation is convenient as one can effectively map any  
    observable to the complex $\Delta$ plane; this allows a straightforward comparison   
    of the sensitivity of the various observables to the effect of the   
    model. Experimental data can be mapped analogously by plotting  
    $\tilde \Delta := O^{Exp}/O^{SM3}$ in the same complex plane. $O^{Exp}$ denotes   
    the experimental value of a generic observable and $O^{SM3}$ the theory prediction  
    within the SM3.  
    In the SM4, we obtain      
      \begin{eqnarray}      
      M_{12}^{K^0, SM4} & \propto & \eta_{cc} \left(\lambda_c^{K^0}\right)^2 S_0(x_c)       
                         + 2 \eta_{ct} \lambda_c^{K^0} \lambda_t^{K^0} S(x_c,x_t)       
                         + \eta_{tt} \left(\lambda_t^{K^0}\right)^2 S_0(x_t)      
      \\ &&      
                         + 2 \eta_{ct'} \lambda_c^{K^0} \lambda_{t'}^{K^0} S(x_c,x_{t'})       
                         + 2 \eta_{tt'} \lambda_t^{K^0} \lambda_{t'}^{K^0} S(x_t,x_{t'})      
                         + \eta_{t't'}  \left(\lambda_{t'}^{K^0}\right)^2 S_0(x_{t'}) \, ,           
       \nonumber \\      
      M_{12}^{B_d, SM4} & \propto & \eta_{tt}   \left(\lambda_t^{B_d}\right)^2 S_0(x_t)       
                                   +\eta_{t't'} \left(\lambda_{t'}^{B_d}\right)^2 S_0(x_{t'})      
                                   + 2 \eta_{tt'} \lambda_t^{B_d} \lambda_{t'}^{B_d} S(x_t,x_{t'})  \, ,     
      \\      
      M_{12}^{B_s, SM4} & \propto & \eta_{tt} \left(\lambda_t^{B_s}\right)^2 S_0(x_t)      
                                   +\eta_{t't'} \left(\lambda_{t'}^{B_s}\right)^2 S_0(x_{t'})      
                                   + 2 \eta_{tt'} \lambda_t^{B_s} \lambda_{t'}^{B_s} S(x_t,x_{t'}) \, .     
        \end{eqnarray}      
     Note that now also those CKM elements change that describe the mixing within the first three families!     
     For simplicity we take the new QCD corrections to be (see also \cite{Soni:2010xh,Buras:2010pi})    
     \begin{equation}     
     \eta_{t't'} = \eta_{tt'} = \eta_{tt} \, \, \, \mbox{and} \, \, \,  \eta_{ct'} =   \eta_{ct} \, .     
     \end{equation}      
     It is interesting to note here that the only information we have currently about the CKM elements $V_{td}$ and    
     $V_{ts}$ comes from $B$-and $K$-mixing plus assuming the unitarity of $V_{CKM3}$.     
     \\   
     \paragraph{The mass difference in the neutral $D$ system}   
can be used to infer  a very strong bound on $|V_{ub'} V_{cb'}|$, see \cite{Golowich:2007ka}\footnote{A similar strategy was   
recently used in \cite{Buras:2010nd}.}. We redid this analysis in   
\cite{Bobrowski:2009ng} and softened the bound.   
     The mass difference in the neutral $D^0$-system is typically  expressed    
     in terms of the parameter $x_D$:      
     \begin{equation}      
     x_D = \frac{\Delta M_D}{\Gamma_D} \leq \frac{2   
           |M_{12}^{D^0}|}{\Gamma_D}\, .      
     \end{equation}      
     For more information on the last inequality see e.g. the discussion in   
     \cite{Bobrowski:2009zc,Bobrowski:2010xg}.   
       HFAG \cite{Barberio:2001mb} quotes for the experimental value of $x_D$   
     \begin{equation}      
     x_D = (0.811 \pm 0.334) \cdot 10^{-2} \, .      
     \end{equation}      
     The main difference compared to the above discussed $K$- and $B$-mixing systems is that in   
     the $D$-system the theory prediction in the SM3 is theoretically not well under control,   
     see e.g. \cite{Bobrowski:2009zc,Bobrowski:2010xg}. However, the pure contribution of a heavy fourth generation   
     to $M_{12}$ can be calculated reliably.   
     Using the unitarity of the CKM matrix of the SM4   
     $\lambda_d^{D^0} + \lambda_s^{D^0} + \lambda_b^{D^0} + \lambda_{b'}^{D^0} = 0$       
     (with $\lambda_{x}^{D^0} = V_{cx} V_{ux}^*$ ),      
     the full expression for $M_{12}$ reads   
     \begin{eqnarray}\label{09022401mb}     
     M_{12}^{D^0} & \propto & \left( \lambda_s^{D^0} \right)^2   S_0(x_s)   
                               + 2 \lambda_s^{D^0} \lambda_b^{D^0}           S(x_s, x_b)   
                              + \left( \lambda_b^{D^0} \right) ^2 S_0(x_b) + \text{LD}   
     \nonumber      
     \\      
     && + 2 \lambda_s^{D^0} \lambda_{b'}^{D^0} S(x_s, x_{b'})+ 2 \lambda_b^{D^0} \lambda_{b'}^{D^0} S(x_b, x_{b'}) + \text{LD}      
     \nonumber      
     \\      
     && + \left( \lambda_{b'}^{D^0} \right)^2 S_0(x_{b'}), \,      
     \end{eqnarray}      
     where the proportionality constant is      
     \begin{equation}     
     \frac{{G_{\text{F}}^2 M_W^2 M_D }}     
     {{12\pi ^2 }}\;f_D^2 B_D \;\eta \left( {m_c ,M_W } \right).     
     \end{equation}     
     We use the same numerical values as in \cite{Bobrowski:2009ng}.   
     The first line of (\ref{09022401mb}) corresponds to the pure SM3 contribution, the third line is due to contributions      
     of the heavy 4th generation and the second line is a term arising when SM3- and $b'$ contributions mix:      
     \begin{eqnarray}      
     M_{12}^{D^0} & = & M_{12,SM3}^{D^0} +  M_{12, Mix}^{D^0} +  M_{12, b'}^{D^0} \, .      
     \end{eqnarray}       
     The idea of \cite{Golowich:2007ka} was to neglect all terms in $M_{12}^{D^0}$, except $ M_{12, b'}^{D^0}$,      
     and to equate this term with the experimental number for $x_D$, since all perturbative short-distance contributions   
     with light internal quarks are negligible.   
     Since it is not completely excluded that there might be large non-perturbative      
     contributions to both $ M_{12,SM3}^{D^0}$  and  $M_{12, Mix}^{D^0}$  (denoted by $LD$),    
     each of the size of the experimental value of $x_D$, we get the following bound   
     \begin{eqnarray}      
     3 M_{12}^{D^0, Exp} & \geq  & M_{12, b'}^{D^0} \, .      
     \end{eqnarray}       
     Allowing this possibility we obtain the following bounds on   
     $|V_{ub'} V_{cb'}|$ 
     \begin{equation}       
     |V_{ub'} V_{cb'}| \leq \left\{      
\begin{array}{cc}      
\label{eq:mass}0.00395 & \, \mbox{for} \, \, \, m_{b'} = 200 \, \mbox{GeV} \, ,      
\\      
0.00290 & \, \mbox{for} \, \, \, m_{b'} = 300 \, \mbox{GeV} \, ,      
\\      
0.00193 & \, \mbox{for} \, \, \, m_{b'} = 500 \, \mbox{GeV} \, .      
\end{array}      
\right.     
\end{equation}      
This bound is still by far the strongest direct constraint on $|V_{ub'} V_{cb'}|$.      
%
%
%
%
%
%
%
%
%
%
%
%
   
\noindent   
\paragraph{Next, we consider the $b\to s \gamma$ transition.} In \cite{Bobrowski:2009ng} we approximated the treatment of the FCNC decay $b \to s \gamma$ by simply looking at the product of CKM structure and the corresponding Inami-Lim function $D_0'(x_t)$    
\cite{Inami:1980fz}\footnote{The Inami-Lim      
     function $D_0'(x_t)$ is proportional to the Wilson-coefficient $C_{7\gamma}(M_W)$.}.
     \begin{equation}      
     \Delta_{b \to s \gamma} :=      
      \frac{|\lambda_t^{SM4}|^2 D_0'(x_t)^2 + 2 \mbox{Re} \left(\lambda_t^{SM4} \lambda_{t'}^{SM4} \right) D_0'(x_t) D_0'(x_{t'})      
            +|\lambda_{t'}^{SM4}|^2 D_0'(x_{t'})^2}     
            {|\lambda_t^{SM3}|^2 D_0'(x_t)^2} \, ,     
     \end{equation}      
     with     
     \begin{equation}     
     D_0'(x) = -\frac{-7x +  5x^2+8x^3}{12(1-x)^3}      
                 + \frac{x^2 (2-3x)}{2(1-x)^4} \ln[x] \,    
     \label{inamilimC7}     
     \end{equation}     
     and $\lambda_x \equiv \lambda_x^{B_s}$.   
     We assumed  that parameters which give a value of $\Delta_{b \to s   
      \gamma}$ close to one will also lead only to small     
     deviations of $\Gamma (b \to s \gamma)^{SM4}/ \Gamma (b \to s   
     \gamma)^{SM3}$ from one. However, this crude treatment imposed   
      a too strong bound on the $3-4$ mixing.   
  
In this work we will use the full leading logarithmic expression for $b \to s \gamma$, see also \cite{Soni:2010xh,Buras:2010pi}. Following \cite{Buchalla:1995vs} we normalize the $b \to s \gamma$ decay rate to the semi leptonic decay rate   
\begin{equation}   
R:= \frac{\Gamma (b \to s \gamma)}{\Gamma ( b \to c e \bar{\nu})}   
  = \frac{|V_{ts}^* V_{tb}|^2}{|V_{cb}|^2} \frac{6 \alpha}{\pi f(z)} \left| C_{7\gamma}^{(0)eff} \right|^2   \, , 
\end{equation}   
$f(z= m_c^2/m_b^2)$ is a phase space factor, which we will not need later on.   
It is interesting to note that in deriving this formula the unitarity of the $3 \times 3$ CKM matrix   
was already used and the CKM combination $\lambda_u = V_{us}^* V_{ub}$ was neglected    
(in comparison to  $\lambda_c$ and  $\lambda_t$).    
The effective Wilson coefficient $ C_{7\gamma}^{(0)eff} $ is a linear combination of the penguin Wilson    
coefficients $C_7$ and $C_8$, which are accompanied by the CKM structure $\lambda_t$   
and the current-current Wilson coefficient $C_2$ with the corresponding  CKM structures $\lambda_c$ and   
$\lambda_u$   
\begin{eqnarray}   
C_{7\gamma}^{(0)eff} (\mu)  & = & C_{7}^{eff1} (\mu)  +   C_{7}^{eff2}  (\mu) +  C_{7}^{eff3}  (\mu)   \, , 
\\   
& C_{7}^{eff1}  (\mu)& = \eta^{\frac{16}{23}} C_{7 \gamma}^{(0)} (M_W)  \, , 
\\   
& C_{7}^{eff2}  (\mu)& = \frac83 \left( \eta^{\frac{14}{23}} -  \eta^{\frac{16}{23}} \right) C_{8 g}^{(0)} (M_W)  \, , 
\\   
& C_{7}^{eff3}  (\mu)& = \sum \limits_{i=1}^8 h_i \eta^{a_i} C_{2}^{(0)} (M_W)  \, , 
\end{eqnarray}   
with   
\begin{equation}   
\eta  (\mu):= \frac{\alpha_s(M_W)}{\alpha_s(\mu)} \, .  
\end{equation}   
The values for $h_i$ and $a_i$ are given in Table XXVII of \cite{Buchalla:1995vs}.   
The initial conditions of the Wilson coefficients read   
\begin{eqnarray}   
C_{2}^{(0)} (M_W) & = & 1  \, , 
\\   
C_{7 \gamma}^{(0)} (M_W) & = & - \frac12  D_0'\left(x_t = \frac{m_t^2}{M_W^2} \right)  \, , 
\\   
C_{8 g}^{(0)} (M_W) & = & - \frac12  E_0' \left(x_t = \frac{m_t^2}{M_W^2} \right)  \, . 
\end{eqnarray}   
$ D_0'(x)$ is given above in Eq. (\ref{inamilimC7}), $ E_0'(x)$ reads   
     \begin{equation}     
    -\frac12 E_0'(x) = -\frac{2x +  5x^2-x^3}{8(1-x)^3}      
                 - \frac{3 x^2 }{4(1-x)^4} \ln[x] \, .     
     \end{equation}     
Numerically it turns out that even for large values of $m_t$ (up to 1000 GeV)  $C_{7}^{eff3}$   
is the dominant contribution to $C_{7\gamma}^{(0)eff} (\mu)$.    
In \cite{Bobrowski:2009ng} we have only taken  $C_{7}^{eff1}$ into account and therefore overestimated the effects of   
a fourth generation to the branching ratio of the decay $b \to s \gamma$.   
Putting everything together we get   
\begin{eqnarray}   
\Delta_{b \to s \gamma} & :=  & \frac{R^\mathrm{SM3}}{R^\mathrm{SM4}}   
\\   
& = &      
\left| \frac{V_{cb}^{\mathrm{SM3}}}{V_{cb}^{\mathrm{SM4}}} \right|^2   
\left| \frac{   
\lambda_c^{\mathrm{SM4}} C_7^{eff3} -\lambda_t^{\mathrm{SM4}} (C_7^{eff1} + C_7^{eff2})  -\lambda_{t'}^{\mathrm{SM4}} (C_7^{eff1'} + C_7^{eff2'})    
             }{   
\lambda_c^{\mathrm{SM3}} C_7^{eff3} -\lambda_t^{\mathrm{SM3}} (C_7^{eff1} + C_7^{eff2})      
} \right|^2  \, . 
\end{eqnarray}   
In \cite{Buras:2010pi} it was suggested to use the LO expression for $b \to s \gamma$ 
at a low scale of $\mu = 3.22$ GeV in order to reproduce the numerical value of the NLO expression. 
We have checked that $\Delta_{b \to s \gamma}$ is quite insensitive to a variation of the scale between $m_b$ and 
3 GeV, so we use $\mu = m_b$.

\subsubsection{FCNC constraints with  sensitivity to lepton mixing}   
   
Next we also discuss FCNC processes that are sensitive to lepton mixing. In principle lepton    
mixing has to be investigated in the same manner as the quark mixing. For simplicity we    
have neglected lepton mixing in this paper. However, we will take into account the    
conservative bounds on $V_{e \nu_4}$, $V_{\mu \nu_4}$ and $V_{\tau \nu_4}$    
given in \cite{Lacker:2010zz} for 
the rare  
decay $B_s \to \mu^+ \mu^-$. 
In the ratio of the SM4 and SM3 predictions   
for the branching ratios almost everything cancels out and one is left with the product of CKM elements and    
Inami-Lim functions   
\begin{eqnarray}   
\Delta_{B_s \to \mu \mu} & := & \frac{BR (B_s \to \mu^+ \mu^-)^{\mathrm{SM4}}}   
                                 {BR (B_s \to \mu^+ \mu^-)^{\mathrm{SM3}}}   
\nonumber 
\\   
& = & \frac{ \left|\lambda_t^{\mathrm{SM4}} Y_0(x_t) +\lambda_{t'}^{\mathrm{SM4}} Y_0(x_{t'})\right|^2}   
           { \left|\lambda_t^{\mathrm{SM3}} Y_0(x_t)                                       \right|^2}  \, , 
\label{DeltaBsmumu} 
\end{eqnarray}   
with the Inami-Lim function 
\begin{equation} 
Y_0[x] = \frac{x}{8} \left( \frac{x-4}{x-1} + 3 \frac{x}{(x-1)^2} \ln [x] \right) \, . 
\end{equation} 
Including also the leptonic contributions we have to make the following substitutions 
\cite{Buras:2010pi} in Eq. (\ref{DeltaBsmumu}) 
\begin{eqnarray} 
Y_0(x_t) & \to &  Y_0(x_t)       - \left|U_{\mu 4}\right|^2 S(x_t, x_{\nu_4}) \, , 
\\ 
Y_0(x_{t'}) & \to & Y_0(x_{t'})  - \left|U_{\mu 4}\right|^2 S(x_{t'}, x_{\nu_4}) \, , 
\end{eqnarray} 
where $S$ is the box function given in Eq. (\ref{box2}), $x_{\nu_4}$ is given by the mass of the fourth neutrino 
and $U_{\mu 4}$ is the PMNS matrix element describing the mixing between the $\mu$ and the fourth neutrino. 
In \cite{Lacker:2010zz} the bound  $U_{\mu 4}< 0.029$ was derived. 
Using this information we find that leptonic contributions give at most a relative correction of 0.5 per mille, 
so we can safely neglect them. 
\\ 
The branching ratio for $B_s \to \mu \mu$ is not measured yet,    
HFAG quotes  \cite{HFAG}  (for the current experimental bound from TeVatron, see also   
\cite{Punzi:2010nv})   
\begin{eqnarray}   
Br(B_s \to \mu^+ \mu^-) & < & 3.6 \cdot 10^{-8}  \, . 
\end{eqnarray}   
In the SM3 one expects a value of    
\cite{Buras:2010pi}   
\begin{eqnarray}   
Br(B_s \to \mu^+ \mu^-) & = & \left( 3.2 \pm 0.2 \right) \cdot 10^{-9}  \, .  
\end{eqnarray}

   
\subsubsection{Allowed ranges for the $\Delta$ parameters}   
\label{allowedDelta}        
     Now we come to a crucial point: the fixing of the allowed ranges for the values   
     of the different $\Delta$s. For our exploratory study - in comparison to a full fit that will  
     be performed in future - we fix reasonable ranges for the $\Delta$s. Therefore we have to investigate  
     theoretical and experimental errors.  
     The FCNC quantities $\Delta M_s$, $\Delta M_d$ and $b \to s \gamma$ are dominated  
     by theoretical uncertainties.  
     Currently, in particular the hadronic uncertainties are under intense   
     discussion, see e.g. \cite{Lenz:2008xt}.   
     Therefore, we use conservative estimates for the theoretical errors. 
      \begin{displaymath}      
      \begin{array}{|c|c|} 
      \hline      
                          &   \hspace{1cm} \mbox{Bound} \hspace{1cm} \, 
      \\      
      \hline \hline      
      |\Delta_{B_d}|      & 1 \pm 0.3       
      \\      
      \hline      
      \phi^\Delta_{B_d}   & 0 \pm 10^\circ  
      \\      
      \hline      
      |\Delta_{B_s}|      & 1 \pm 0.3       
      \\      
      \hline      
      \phi^\Delta_{B_s}   & \mbox{free}     
      \\      
      \hline      
      \mbox{Re} (\Delta_{K})      & 1 \pm 0.5  
      \\      
      \hline      
      \mbox{Im} (\Delta_{K})      & 0 \pm 0.3  
      \\      
      \hline      
      \Delta_{b \to s \gamma}      & 1 \pm 0.15 
      \\      
      \hline      
      \Delta_{B_s \to \mu \mu}      & <15       
      \\      
      \hline      
       \end{array}      
      \end{displaymath}      
Since we choose for the central values of our $\Delta$s the value one, all resulting    
allowed parameter points for $V_{CKM4}$ will be scattered around the SM3 values by definition.   
For a future fit we will use $\Delta$s with the central value $\tilde \Delta =  O^{Exp}/O^{SM3}$.  
\\   
This means in other words that we do not take into account some current deviations  
in flavor physics in our current analysis, we simply include them in our error band for the $\Delta$s. 
\section{Putting things together---Constraints on the parameter space}     
In order to   
constrain the mixing   
with the fourth quark family we perform a scan through the parameter space of the model. To this end   
we use the exact parameterization of $V_{CKM4}$ described   
in Sec. 2.1, Eq. \eqref{eq:CKM4FP}.   
For the tree-level bounds we use the central values and standard deviations   
as given in \ref{sec:tree}; we allow for a variation at the  $2\sigma$ level. 
The restrictive Peskin-Takeuchi parameters are allowed to vary at the $99\%$ 
confidence level of \cite{Erler:2010sk}, cp.~Sec.~\ref{STUFit}. 
For the quark masses we use a hard lower limit of   
$280 \; \rm GeV$ and allow for a maximal mass difference of $80\;\rm GeV$   
as  determined in Sec.~\ref{ewsect}. The lepton masses are chosen   
to be larger than $100 \; \rm GeV$ with a maximal splitting of $140\;\rm GeV$.  
For the FCNC we use the $\Delta$s given in Sec \ref{allowedDelta}. 
\\ 
Then we generate a large number ($\mathcal{O}(10^{11})$ ) of randomly   
distributed points in the 13 dimensional parameter   
space\footnote{$m_{t'}, m_{b'}, m_{l_4}, m_{\nu_4}, \theta_{12}, \theta_{23}, \theta_{13},  
\theta_{14}, \theta_{24}, \theta_{34}, \delta_{13},\delta_{14}, \delta_{24}$.}.   
For each point we determine the value of CKM matrix elements,   
flavor and electroweak observables and check whether the various   
experimental bounds are passed (for details see \cite{Bobrowski:2009ng}).

   
\subsection{Result for the mixing angles}   
   
Let us for the moment ignore correlations among the various parameters and focus on  
the maximally allowed size of the mixing with the fourth generation. Table \ref{AllowedMixing}  
shows the limits on the mixing angles $\theta_{14}$,  $\theta_{24}$, and $\theta_{34}$ --- without and with the electroweak bounds.   
 \begin{table}\begin{center}  
\begin{tabular}{|c||c|c|c|}   
 \hline  
&only tree-level & with FCNC bounds & with electroweak  observables\\ \hline \hline  
  $\theta_{14}$  & $<0.07$ &  $<0.0535$  & $< 0.0535$  \\ \hline  
  $\theta_{24}$  & $<0.19$ &  $<0.145 $  & $< 0.121 $  \\ \hline  
  $\theta_{34}$  & $<0.8 $ &  $<0.67  $  & $< 0.35 $  \\ \hline  
\end{tabular}  
 \caption{Maximal mixing of SM fermion generations with the fourth generation.  
The left column show the effect of tree-level bounds alone,   
for the central column FCNC bounds where added and the right-most column gives the  limits including electroweak parameters. \label{AllowedMixing}}  
\end{center}  
\end{table}  
  
\noindent       
As already expected in Sec.~2, the Peskin-Takeuchi parameters
impose strong constraints on the mixing of Standard Model
fermions with the fourth generation.
These numbers are comparable with the ones quoted  
in \cite{Chanowitz:2009mz,Buras:2010pi}. 
\noindent 
The most dramatic effect is observed for the mixing angle 
$\theta_{34}$. The maximal size is roughly halfed by the
virtue of the $T$ parameter alone. So, already at this stage, one is able to conclude that a study of the flavor aspects of SM4 must
not be decoupled from a simultaneous analysis of the 
electroweak sector.
\\  
To illustrate the dependence of $T$ on $\theta_{34}$ we also show the scatter plot for $T$ versus $\theta_{34}$:  
\begin{center}  
 \begin{center}  
 \includegraphics[width=0.4\textwidth]{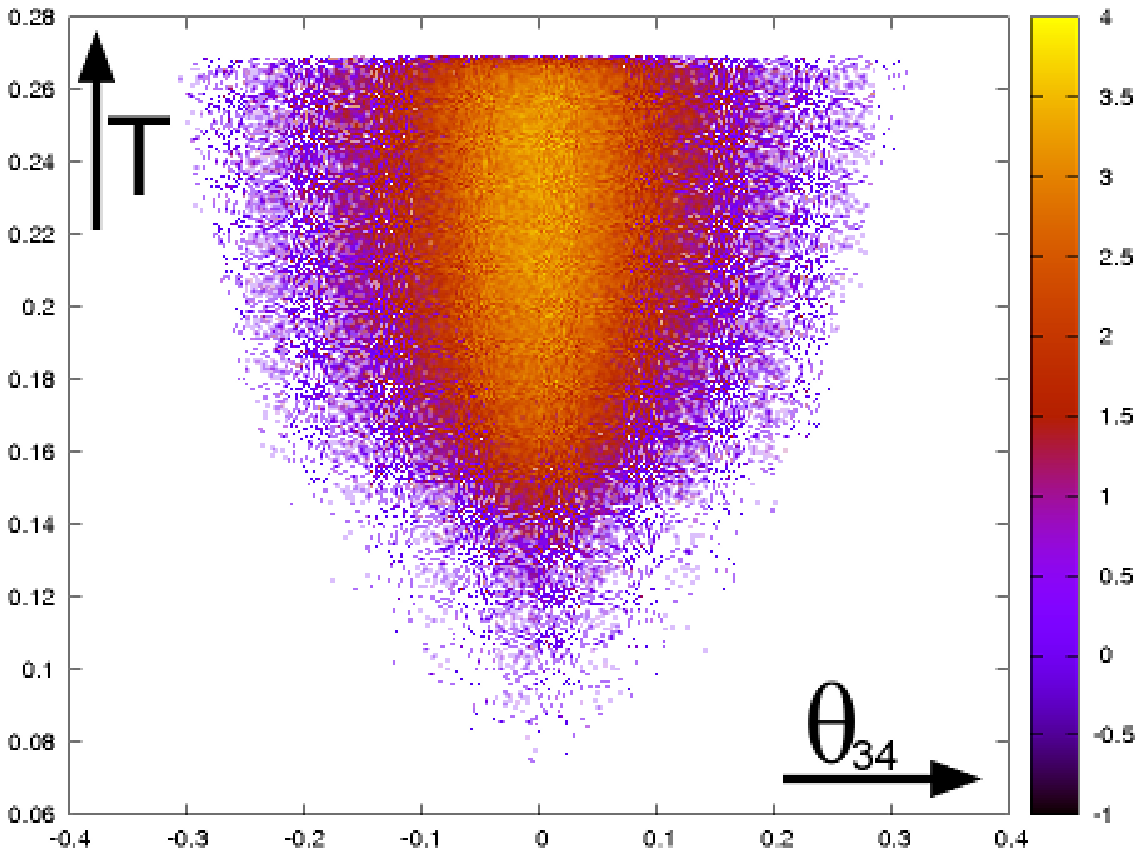}  
\end{center}  
\end{center}

\begin{figure}  
 \begin{minipage}{0.32\textwidth}  
  \begin{center}  
 \includegraphics[width=0.98\textwidth]{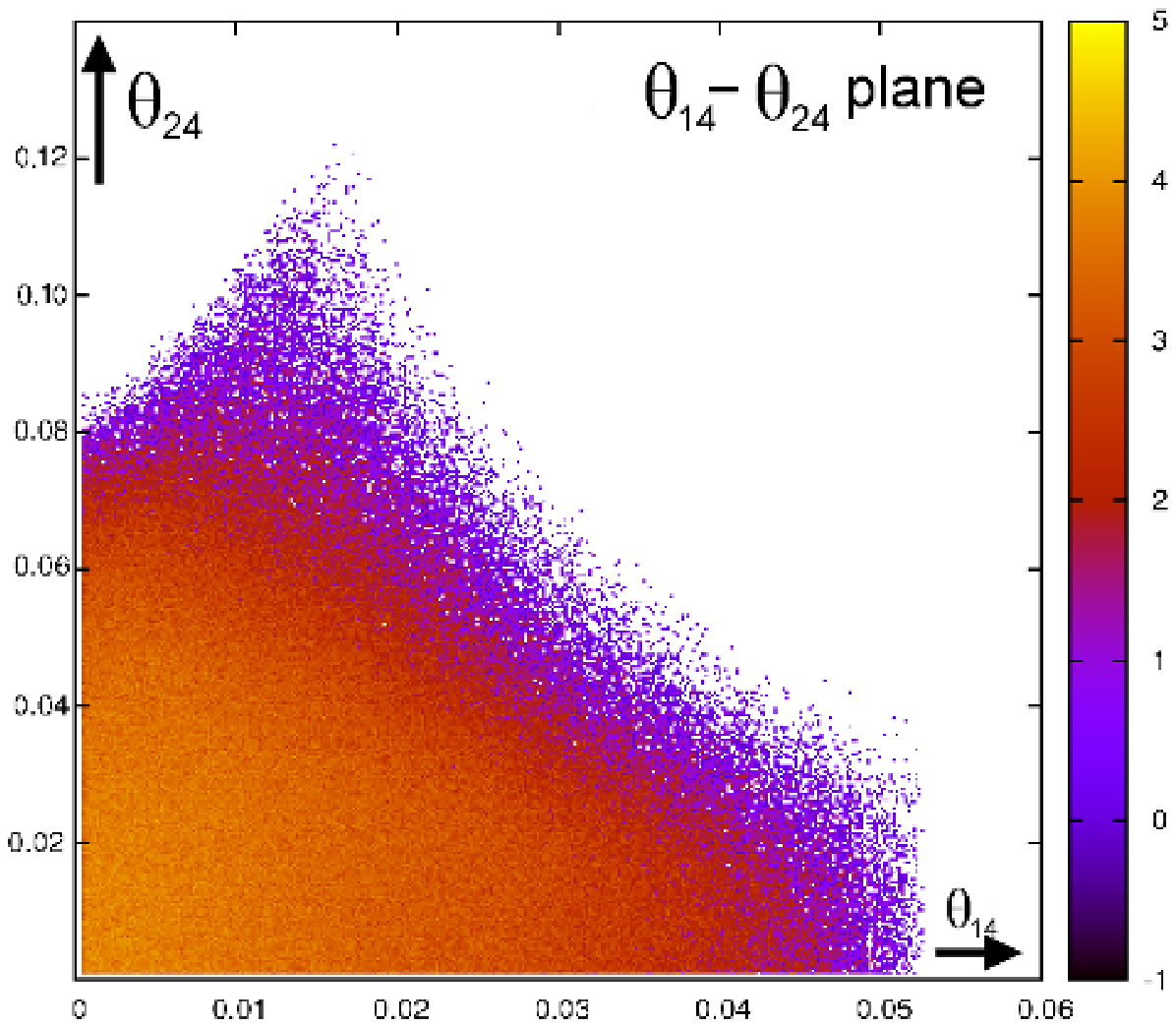}  
\end{center}  
 \end{minipage}$\;\;$  
 \begin{minipage}{0.32\textwidth}  
    \begin{center}  
 \includegraphics[width=0.98\textwidth]{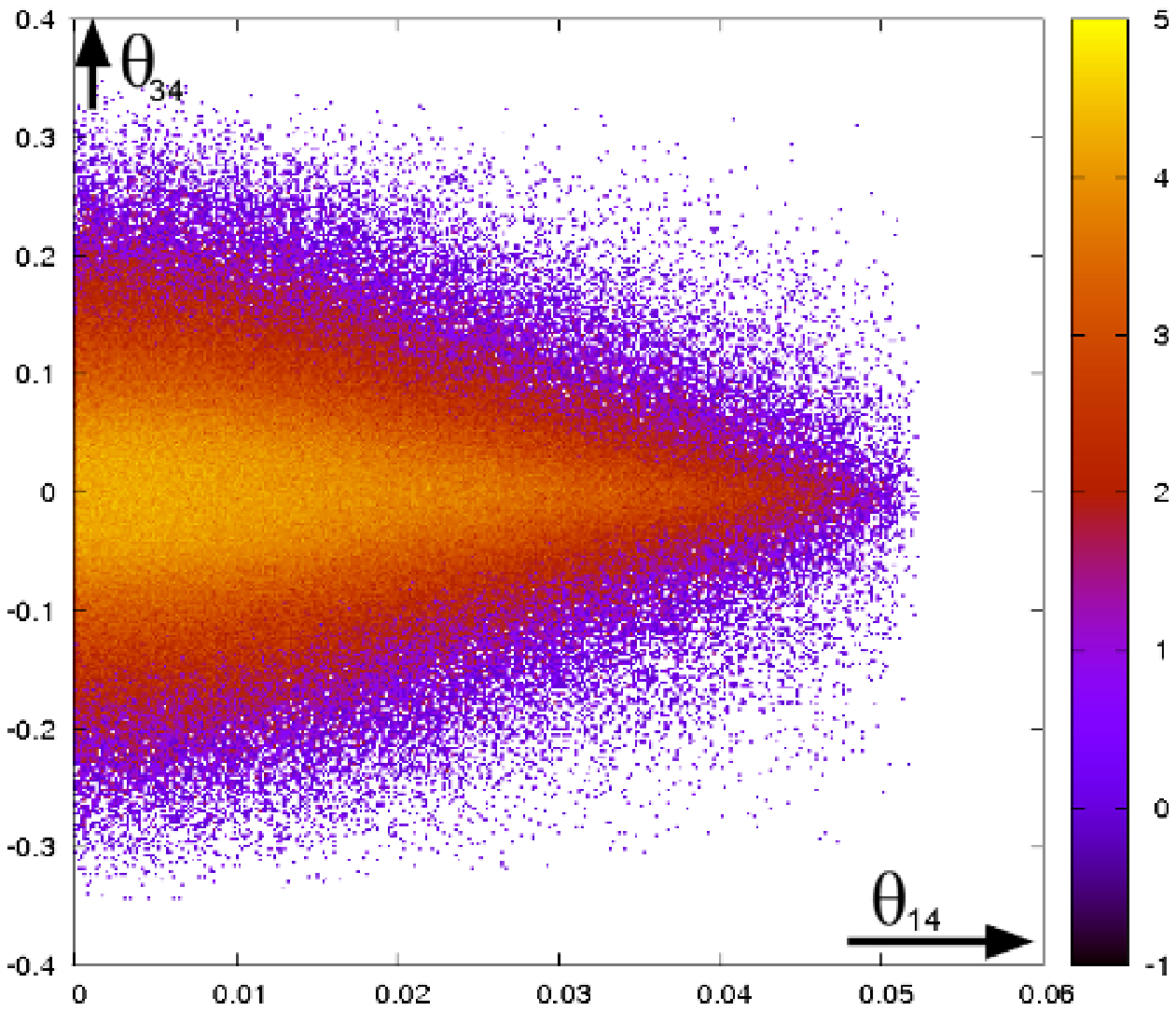}  
\end{center}  
 \end{minipage}$\;\;$  
 \begin{minipage}{0.32\textwidth}  
      \begin{center}  
 \includegraphics[width=0.98\textwidth]{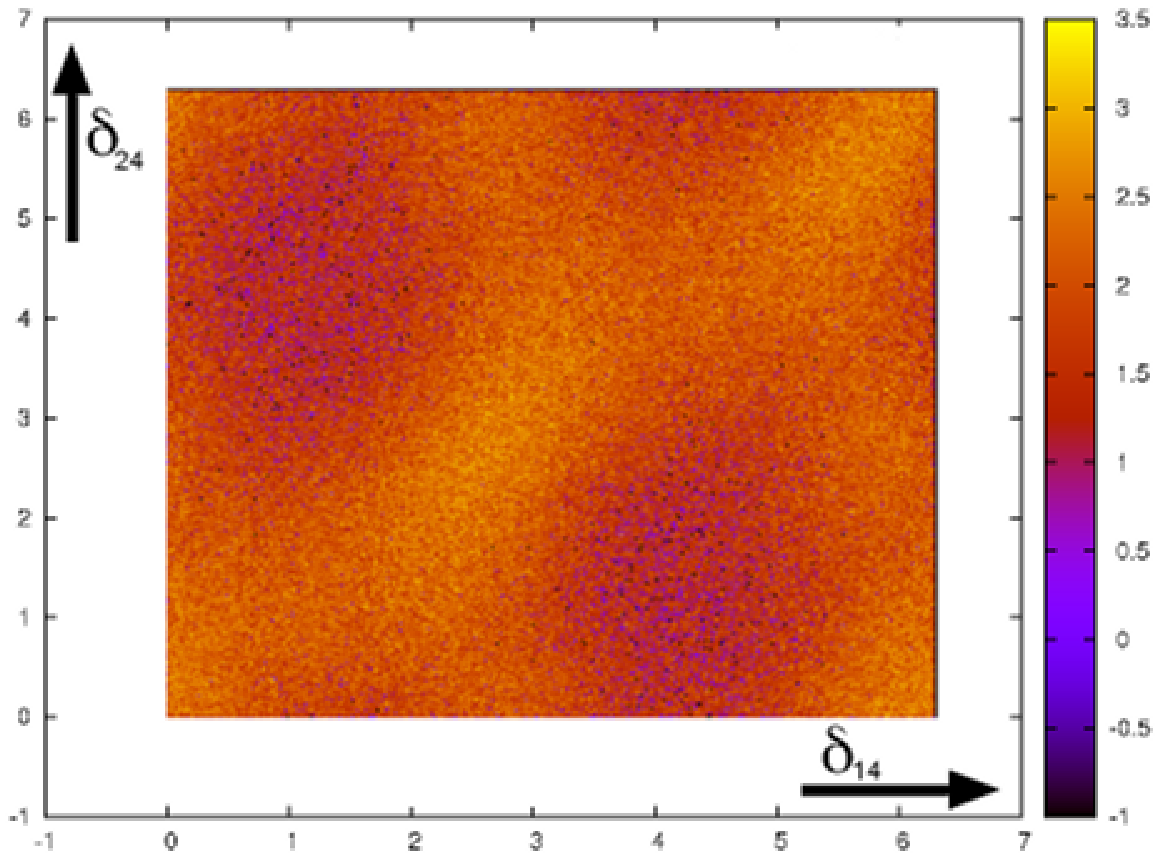}  
\end{center}  
 \end{minipage}  \\
 \begin{minipage}{0.32\textwidth}  
  \begin{center}  
 \includegraphics[width=0.98\textwidth]{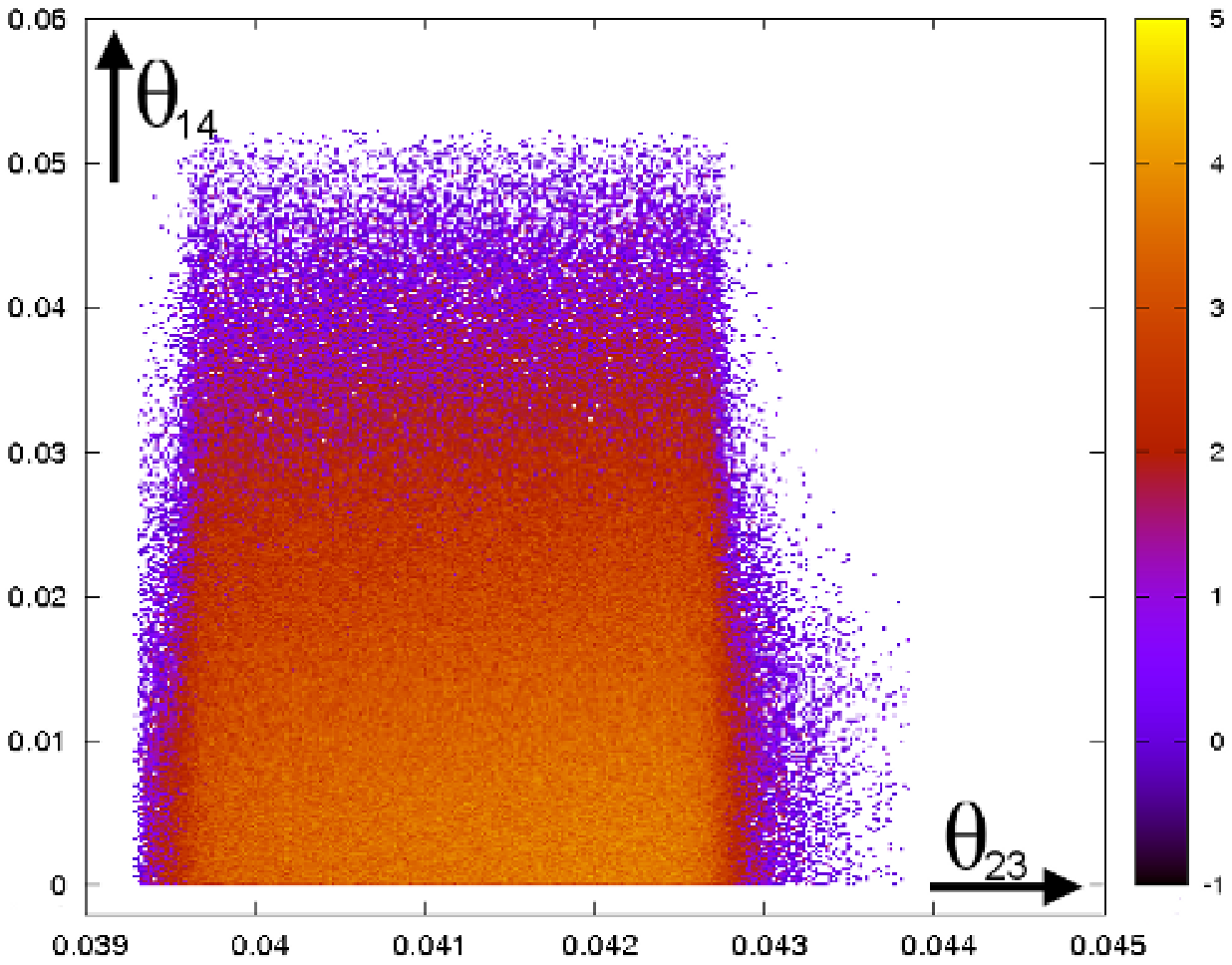}  
\end{center}  
 \end{minipage}$\;\;$  
 \begin{minipage}{0.32\textwidth}  
    \begin{center}  
 \includegraphics[width=1.0\textwidth]{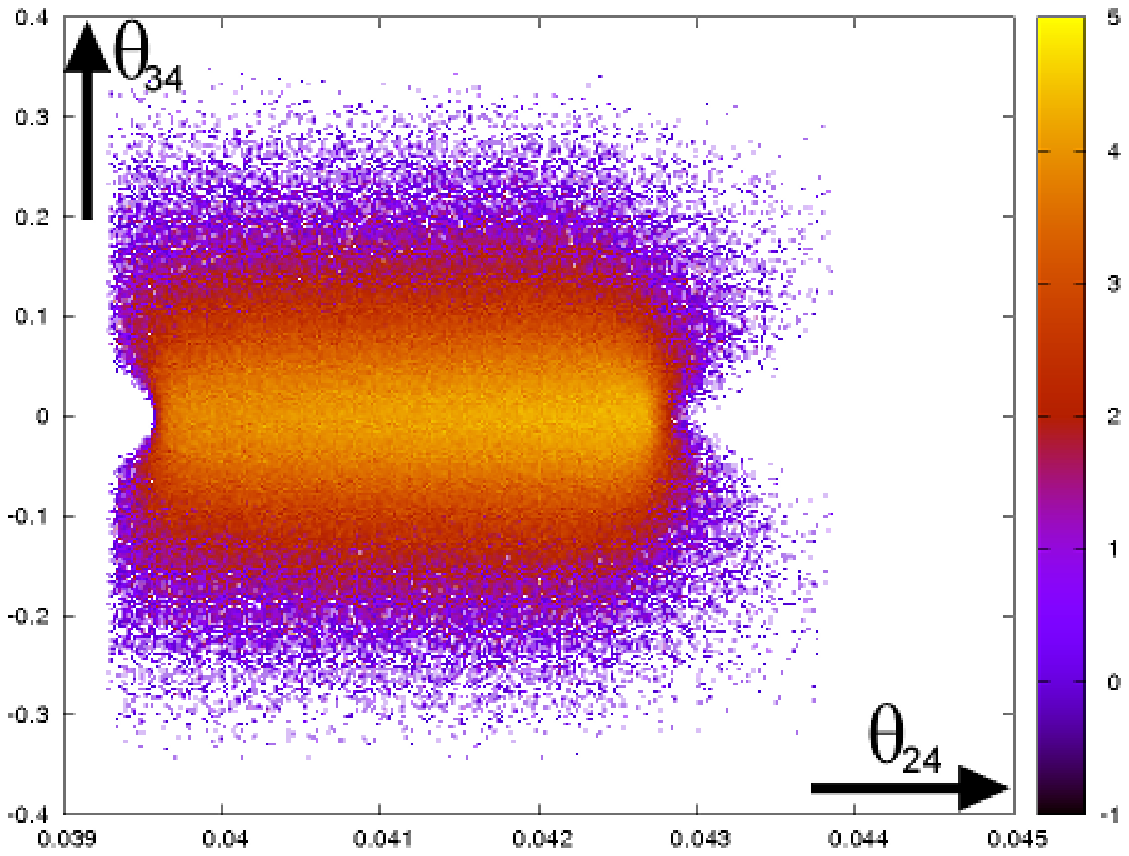}  
\end{center}  
 \end{minipage}$\;\;$  
 \begin{minipage}{0.32\textwidth}  
      \begin{center}  
 \includegraphics[width=0.9\textwidth]{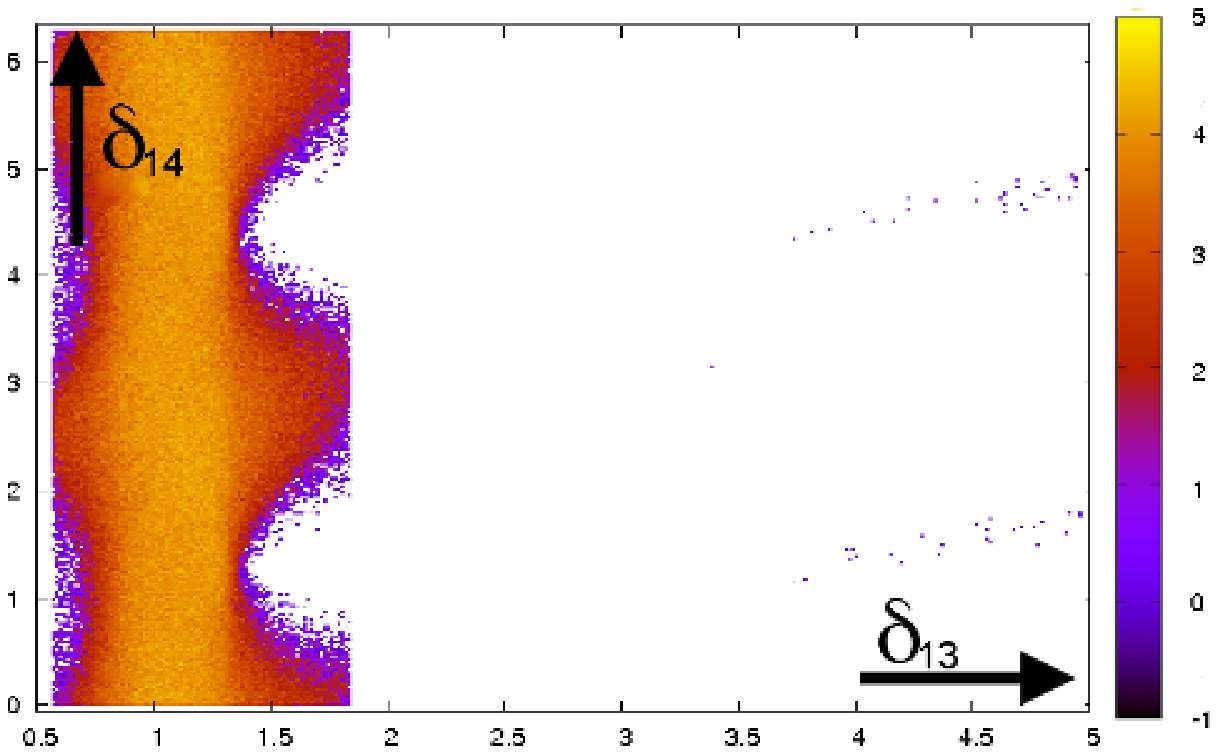}  
\end{center}  
 \end{minipage}  
\caption{Some correlations of the angles and phases. \label{AngleCorrelations}}  
\end{figure}

\noindent  
Next, the correlations between the different angles are examined.  
The results are depicted in Fig.~\ref{AngleCorrelations}.  
Obviously, the maximal mixing angles given in Table~\ref{AllowedMixing}  
cannot be simultaneously realized; especially $\theta_{14}$ and $\theta_{24}$  
show a rather strong correlation and maximal $\theta_{24}$ is only possible  
for $\theta_{14}$ close to $0.018$. Note e.g. ~the $\theta_{14}-\theta_{34}$ correlation; 
simultaneous large mixing angles $\theta_{34}$ and $\theta_{14}$  are also disfavored. 
Indeed, this observation is rather natural, as $V_{td}$ includes a term 
$c_{12} c_{13} c_{24} s_{14}s_{34} e^{i \delta_{14}}$. Hence, simultaneous  
large $s_{14}$ and $s_{34}$ would lead to  a large modification of $V_{td}$; $
B_d$ mixing would be sensitive to this and indeed proves to be the most restrictive of the mixing observables. 
Due to this observation we disfavor a strategy based on starting from fixed bounds on the 
mixing angles without taking the correlations into account.
\\ 
The large 'voids' in the $ \delta_{13}-\delta_{14}$ plane
can be traced to the effect of the direct limit on the phases due to the CKM angle $\gamma$.

\subsection{Results for the CKM elements}   
   
\begin{figure}  
 \begin{minipage}{0.32\textwidth}  
  \begin{center}  
 \includegraphics[width=0.98\textwidth]{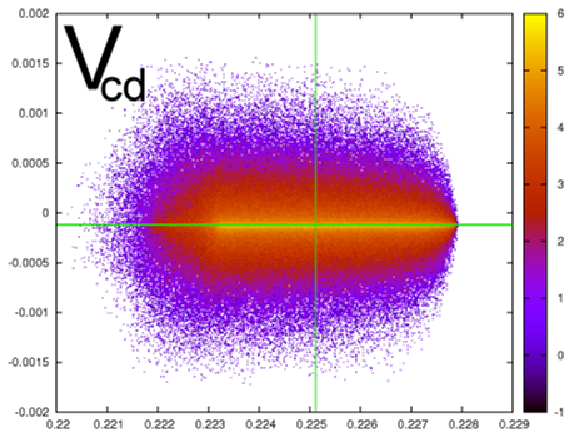}  
\end{center}  
 \end{minipage}$\;\;$  
 \begin{minipage}{0.32\textwidth}  
    \begin{center}  
 \includegraphics[width=0.98\textwidth]{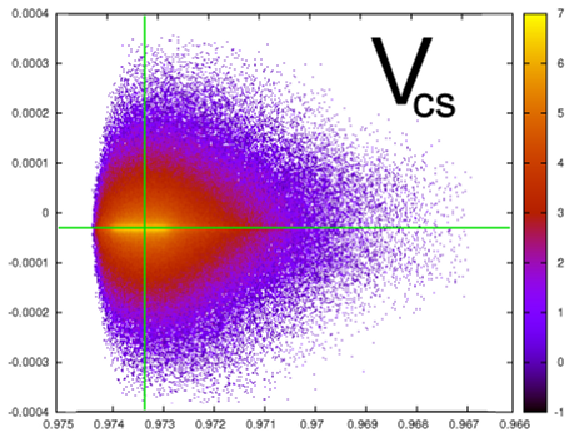}  
\end{center}  
 \end{minipage}$\;\;$  
 \begin{minipage}{0.32\textwidth}  
      \begin{center}  
 \includegraphics[width=0.98\textwidth]{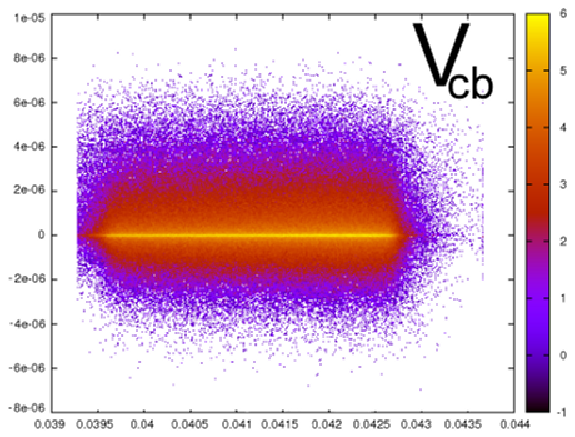}  
\end{center}  
 \end{minipage}\\ 
 \begin{minipage}{0.32\textwidth}  
  \begin{center}  
 \includegraphics[width=0.98\textwidth]{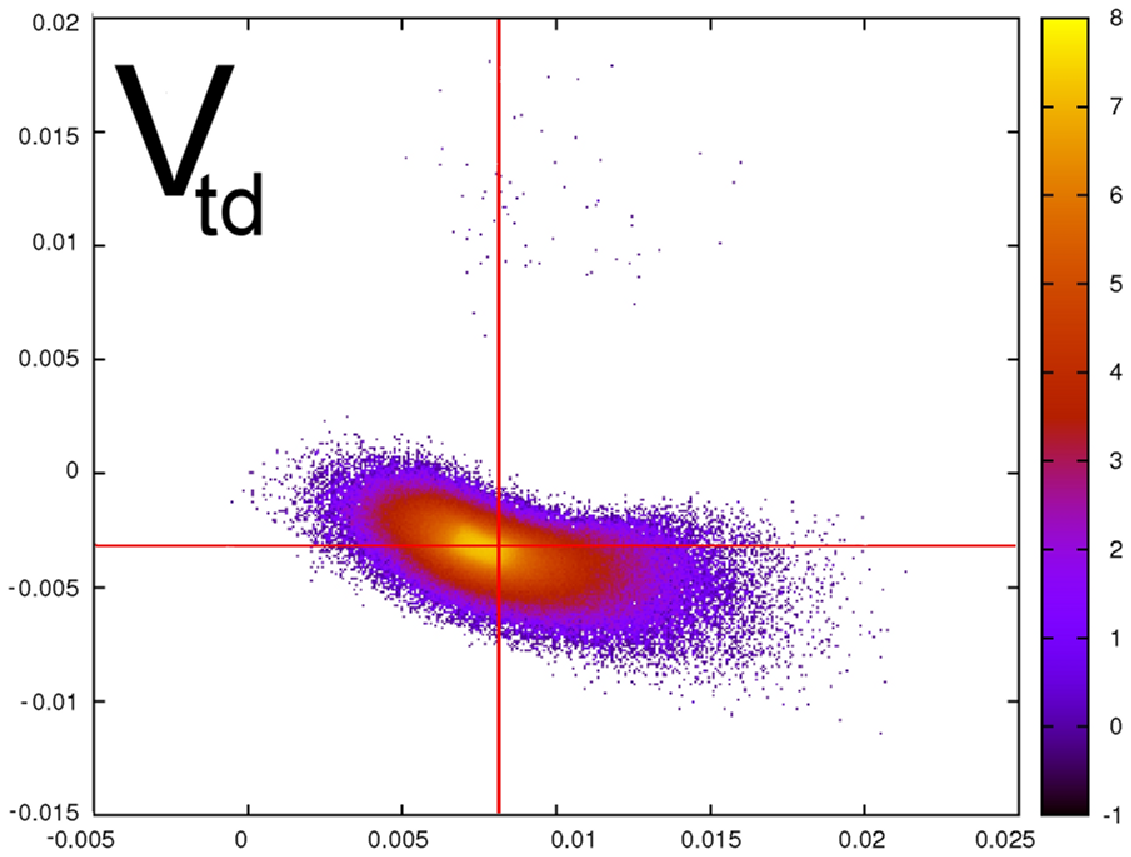}  
\end{center}  
 \end{minipage}$\;\;$  
 \begin{minipage}{0.32\textwidth}  
    \begin{center}  
 \includegraphics[width=0.98\textwidth]{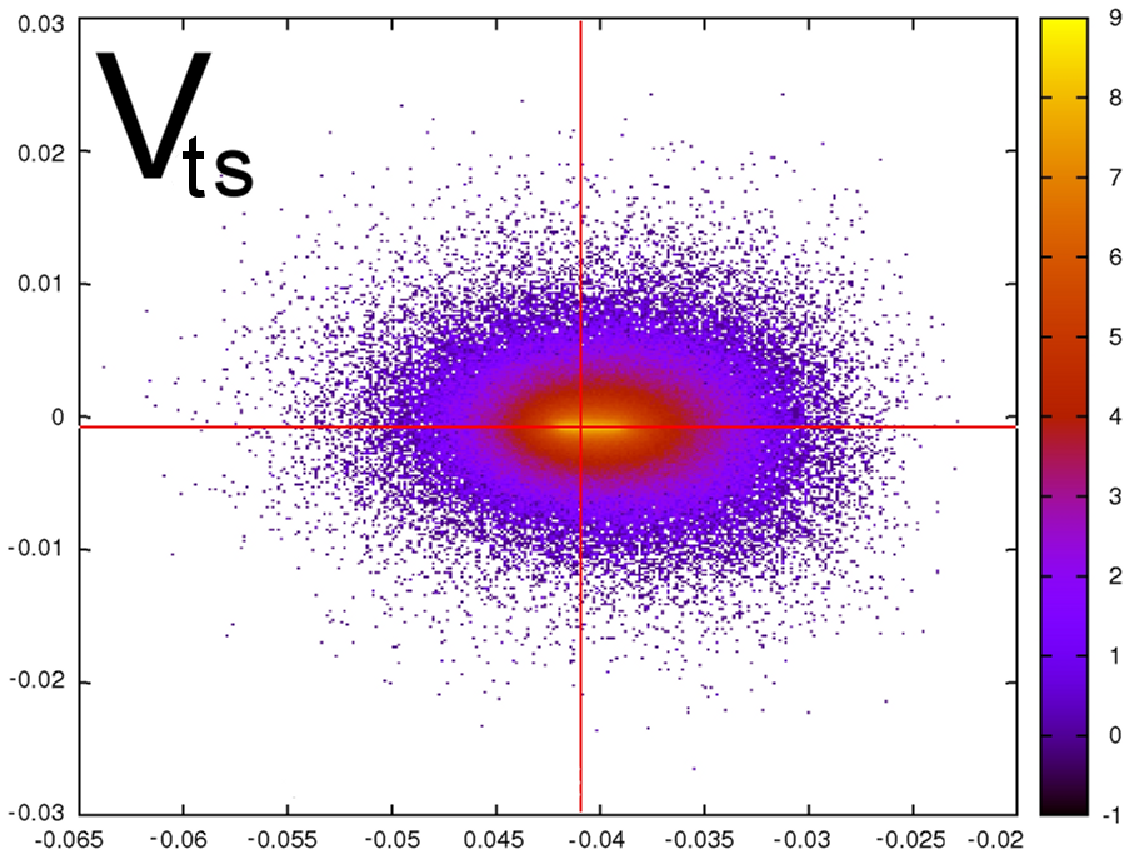}  
\end{center}  
 \end{minipage}$\;\;$  
 \begin{minipage}{0.32\textwidth}  
      \begin{center}  
 \includegraphics[width=0.98\textwidth]{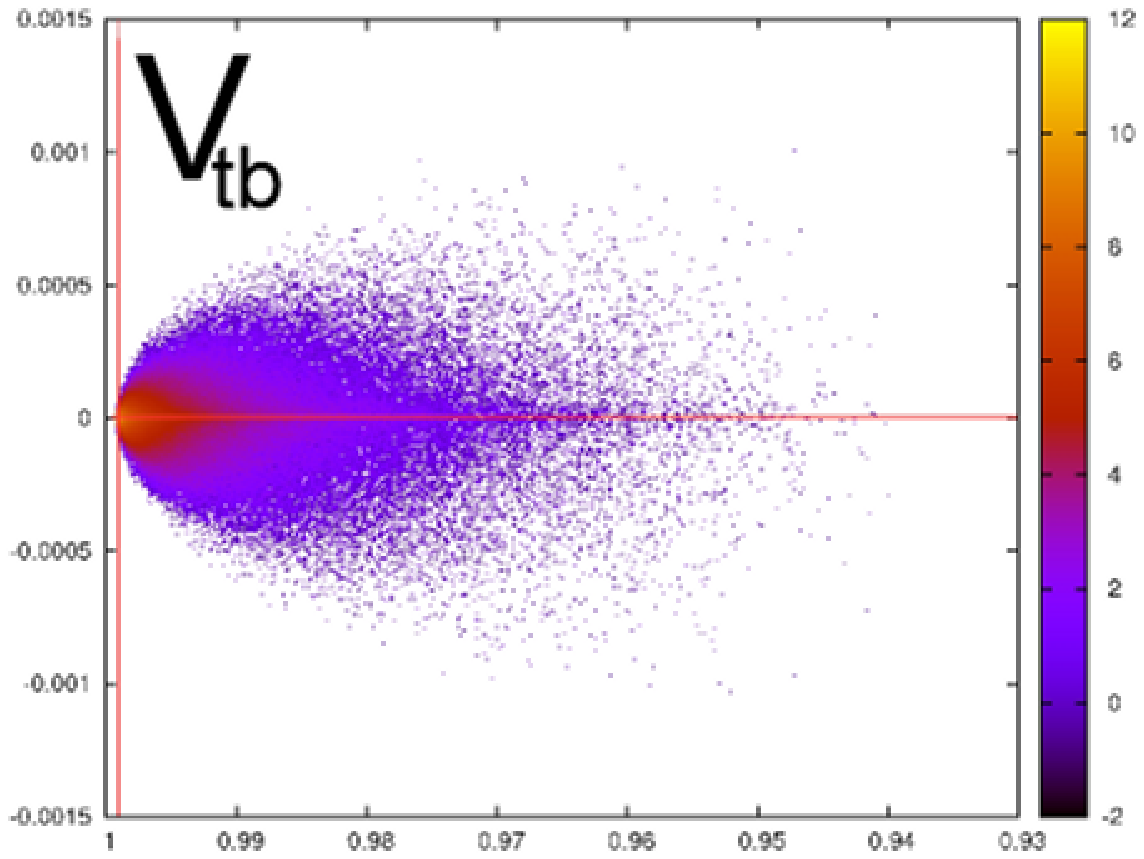}  
\end{center}  
 \end{minipage}  
\caption{Possible modifications of the SM3 CKM matrix elements  
 $V_{cd}$, $V_{cs}$, $V_{cb}$ $V_{td}$, $V_{ts}$ and $V_{tb}$ in  
the SM4 scenario. Depicted is the real part versus the imaginary 
 part of the CKM element (in the standard representation). The crossed  
lines show the SM3 value; $V_{cb}$ is real by construction in the SM3.  \label{CKMelements}}  
\end{figure}

Since the $3\times 3$ unitarity fixes the values of the second and third row CKM elements  
rather precisely in the SM3, it is interesting to see the effect of the   
lifting of the unitarity constraint on $V_{cd}$, $V_{cs}$,  
$V_{cb}$, $V_{td}$, $V_{ts}$ and $V_{tb}$.  
In Figure  \ref{CKMelements} we present the  possible values for these CKM   
matrix elements in the complex plane. Note   
that a CKM matrix element itself is not a physical observable as it depends   
on phase conventions and CKM parameterization; one can, however, compare   
the values for the elements once the representation and phase convention is fixed.   
The plots correspond to the standard represention, cp.~\eqref{CKM3},  
which is the limit of the Botella-Chau representation for zero mixing with a fourth family. 
\\ 
The absolute value of the elements of the second row cannot change  
much with respect to the SM3; however, it is interesting to observe that  
the imaginary part of $V_{cd}$ and $V_{cs}$ can be increased by an order 
of magnitude. This might be potentially interesting for searches for CP violation  
in the charm sector. While $V_{cb}$ does not have an imaginary part in the  
standard representation in SM3, a tiny imaginary part can be present in  
SM4. 
\\  
The absolute value of both, $V_{td}$ and $V_{ts}$, can be modified   
(with respect to their SM3 value) by approximately a factor $2$.  
More important, the imaginary part of $V_{ts}$ can be an order of   
magnitude larger than in the SM. Therefore, one can expect that  
the weak phases of processes involving $V_{ts}$, e.g.~ $B_s$ mixing,  
may experience large corrections.  
\\  
The absolute value of $V_{tb}$ can be as low as $0.93$. Without   
the constraints coming from oblique parameters this limit would   
be  much lower --- around $0.8$. This again shows that   
the electroweak sector imposes strong limits on the flavor   
structure of SM4.   
\\ 
This number is in particular interesting since it can be compared with direct  
determinations of $V_{tb}$ from Single-top-production from TeVatron 
\cite{Abazov:2009ii,Aaltonen:2009jj,Group:2009qk} 
\begin{equation} 
V_{tb}^{\tt TeVatron} = 0.88 \pm 0.07 \, . 
\end{equation}

\subsection{New physics in $B_s$-mixing}   
  
\begin{figure}  
\begin{minipage}{0.47\textwidth}     
\begin{center}    
\includegraphics[width=0.98\textwidth]{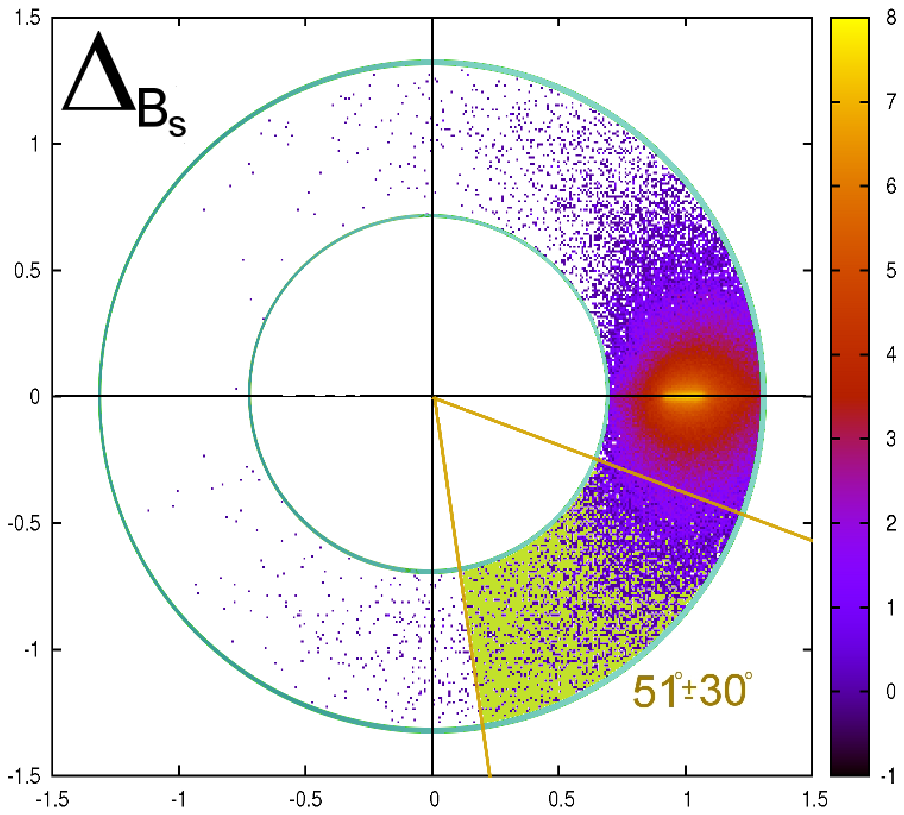}   
\end{center}    
\end{minipage}  
$\;$  
\begin{minipage}{0.47\textwidth}     
\begin{center}    
\includegraphics[width=0.98\textwidth]{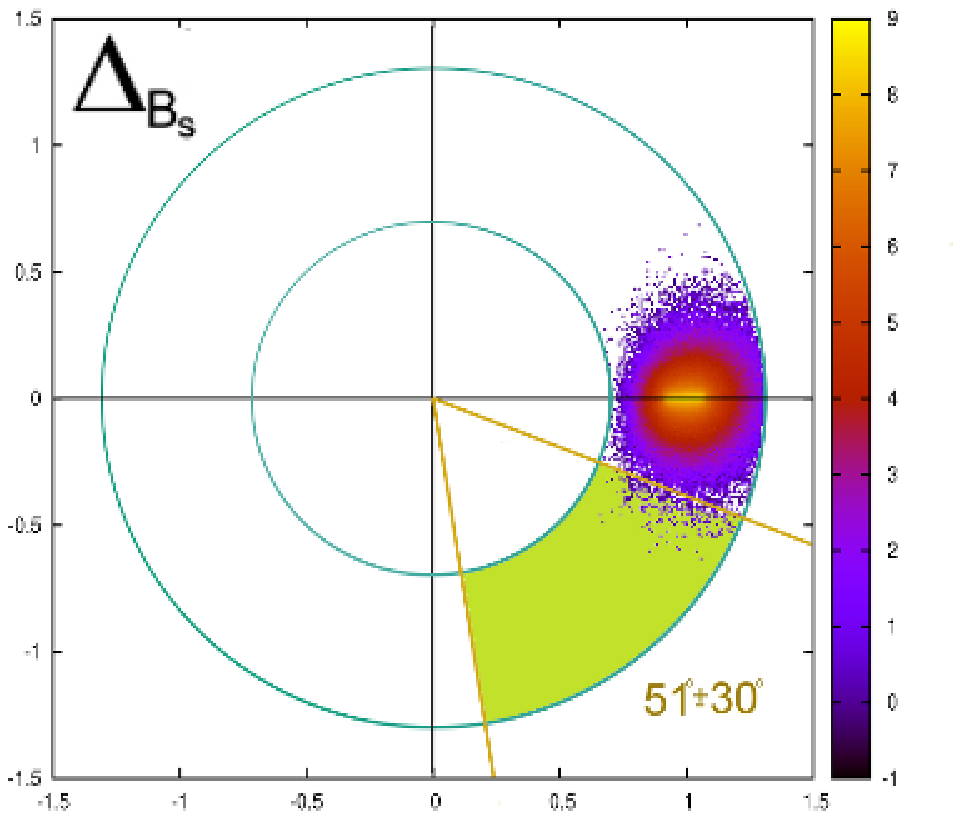}  
\end{center}    
\end{minipage}  
\caption{Complex $\Delta$ plane for $B_s$ mixing.   
The green shaded area corresponds roughly to the experiment at the 1$\sigma$ level \cite{HFAG}.  
The left  panel shows the possible phase if one omits the $T$ parameter,   
the right panel shows the impact of $T$ on the allowed  
phase $\Phi_s$. \label{fig:Bs}}  
\end{figure}

The results for the complex $\Delta_{B_s}$ plane is particularly interesting since there might be some     
hints on new physics effects in the CP-violating phase of $B_s$ mixing, see \cite{Lenz:2006hd,Tarantino:2009sx}     
and the web-updates of \cite{Hocker:2001xe}.     
In \cite{Lenz:2006hd} a visualization of the combination of the mixing quantities $\Delta M_s$,     
$\Delta \Gamma_s$, $a_{sl}^s$,   which are known to NLO-QCD     
\cite{Buras:1990fn,Beneke:1998sy,Beneke:2003az,Ciuchini:2003ww}      
and of direct determinations of $\Phi_s$ in the complex $\Delta$-plane was suggested.     
Combining recent measurements \cite{HFAG, Tevatron} for the phase $\Phi_s$     
one obtains a deviation from the tiny SM-prediction \cite{Lenz:2006hd} in the range     
of 2 to 3 $\sigma$:    
\begin{itemize}    
\item  HFAG:       2.2 $\sigma$ \cite{HFAG},     
\item  CKMfitter: 2.1...2.5 $\sigma$ \cite{Deschamps:2008de, Charles},     
\item  UTfit:     2.9 $\sigma$ \cite{Tarantino:2009sx}.     
\end{itemize}    
The central values of these deviations cluster around    
\begin{equation}     
\Phi_s \approx - 51^\circ.     
\end{equation}     
Very recently the D$0$ collaboration announced a $3.2 \sigma$ deviation of a linear combination 
of $a_{sl}^d$ and $a_{sl}^s$ \cite{Abazov:2010hv} 
from the standard model prediction in \cite{Lenz:2006hd}. This deviation also 
indicates a large negative value of $\Phi_s$. 
\\ 
As can be read off from the left picture of Figure \ref{fig:Bs} sizeable values for $\Phi_s$ can also be   
obtained in scenarios with additional fermions.     
Such large values for $\Phi_s$ are not favored, but they are possible. An    
enhancement of $\Phi_s$ to large negative values by contributions of a fourth generation     
was first predicted in \cite{Hou:2006mx}, by choosing the parameters of the fourth generations  
in such a way that other flavor problems like the $B \to K \pi$ puzzle are solved.  
%
Once the T parameter is implemented with full CKM dependence and used as an additional bound on the CKM elements,   
a large value of $\Phi_s$ seems to be very unlikely and requires a significant fine tuning of the parameters,  
see the right picture of Figure \ref{fig:Bs}.  
In that respect we differ slightly from the conclusion of e.g.  
\cite{Hou:2006mx,Soni:2010xh,Buras:2010pi,Hou:2010mm}, where very large values for $\Phi_s$ are  
allowed\footnote{If we also described the problems in e.g. $B \to K \pi$ by a fourth family, we also 
would exclude the points with $\Phi_s$ close to zero and we would predict a sizeable phase - around $-20^\circ$ - 
but no points with e.g. $-50^\circ$ survive in our analysis. Because of hadronic uncertainties we did not include 
$B \to K \pi$ in our analysis.}. 
\\ 
Here we expect new and considerably more precise data from TeVatron and LHC soon. If the central value  
stayed at the current position, the possibility would arise to find new physics that can not originate  
from an additional fermion family alone.  
 
\begin{figure} 
 \begin{center} 
 \includegraphics[width=0.7\textwidth]{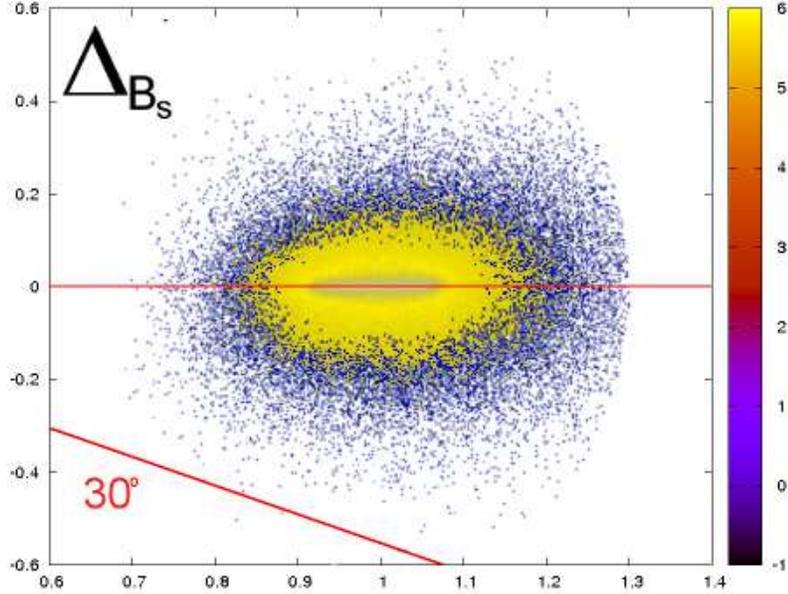} 
\end{center} 
\caption{Dependence of $\Delta_{B_s}$ on the $t'$ mass. The dark blue points  
correspond to masses below $440\;\rm GeV$, the yellow (light grey) point correspond  
to masses heavier than $440\;\rm GeV$. The red line indicates a phase $\Phi_s$ of  
$30^\circ$.\label{BsMass}} 
\end{figure}

It is interesting to study the mass dependence of the phase $\Phi_s$. 
Already in \cite{Hou:2010mm} it was noted that large phases clearly favor 
small $t'$ (and $b'$) masses and the largest phases require a value of $m_{t'}$ 
close to $300\;\rm GeV$; this behaviour is also present in our analysis 
and it seems that too large quark masses struggle to resolve the  
tension in the flavor sector should future experiments confirm e.g.~a  
phase of the order of $30^\circ$. In Fig.\ref{BsMass} we show the distribution 
of the scatter plot for ``light'' ($<440\;\rm GeV$) and ``heavy''($>440\;\rm GeV$)  
$t'$ masses\footnote{Note that the points corresponding to ``heavy'' $t'$ quarks are  
placed on top of the one corresponding to ``light'' $t'$s. A light $t'$ 
and a simultaneous small phase $\Phi_s$ are, of course, still possible. }. 
 
As a final remark, we would like to point out that  
the naively expected huge contributions to $\Delta_{B_s}$ due to the  
heavy $t'$ can (still) be veiled by the mechanism described in  
\cite{Bobrowski:2009ng}. The new contributions in the SM4 fall into  
two classes. The first class contains the 'direct' effects of the 
new heavy ferminons, i.e., the contributions of diagrams with  
at least one $t'$ in the loop. The second class is more subtle 
as it includes the 'indirect' effects of the SM4 scenario. These 
contributions arise from the breaking of the $3 \times 3$ unitarity; 
the standard model CKM elements have to be 'thinned' 
out to accomodate for the non-zero values of the fourth row and column  
matrix elements: this results in a modification of the SM-like contributions. 
 
These two sets of contributions are both sizeable (typically large enough to  
violate at least one experimantal bound), but they can cancel to a very large extent. 
Here is one example for the case of $\Delta_{B_s}$ which survived all our constraints:\\ 
\begin{center} 
{\small 
\begin{tabular}{|c|c|c|c|c|c|c|c|c|c|c|c|c|}\hline 
  $\theta_{12}$ & $\theta_{13}$ & $\theta_{23}$ & $\theta_{14}$ & $\theta_{24}$ &  
  $\theta_{34}$ & \\ \hline 
  $0.2275$ & $0.003409$ & $0.04036$ & $0.01701$ & $0.08392$ &  
  $0.1457$& \\ \hline\hline 
  $\delta_{13}$ &  $\delta_{14}$ &   $\delta_{24}$ &   
  $m_{t'}$ & $m_{b'}$ & $m_{l_4}$ & $m_{\nu_4}$ \\ \hline 
   $1.019$ &  $0.91825$ &   $1.0787$ &   
  $385\;\rm GeV$ & $378\;\rm GeV$ & $602\;\rm GeV$ & $636\;\rm GeV$\\ \hline 
\end{tabular}} 
\end{center} 
This parameter set yields 
\begin{align} 
 \Delta_{B_s}=1 + \underbrace{0.5379 - 0.9016 i }_{\rm direct}  
              \underbrace{-0.4196 + 0.4262 i}_{\rm indirect} = 1.2\; e^{-i 23^\circ} \, . 
\end{align} 
One sees that each contribution is of the size of the Standard model result; 
however, direct and indirect contribution have opposite signs and cancel  
to a large extent. Note that in this case the 3-4 mixing is strong enough  
to allow for almost degenerate quark masses. 
\section{Taylor expansion of $V_{CKM4}$}      
\setcounter{equation}{0}      
      
In \cite{Bobrowski:2009ng} we gave a Wolfenstein-like expansion of the CKM matrix  
for four fermion generations, which allows for  a first estimate of  possible effects  
of the fourth generation. Since the inclusion of the electroweak oblique parameters led  
to tighter constraints for the mixing, one can now further improve this expansion.   
\\  
In the SM3 the hierarchy of the mixing between the three quark families can  
conveniently be visualized via a Taylor expansion in the small CKM element  
$V_{us} \approx 0.2255=\lambda$: the Wolfenstein parameterization \cite{Wolfenstein:1983yz}.    
\\   
Following \cite{Buras:1994ec}\footnote{Note that due to historical reasons the element  
$V_{ub}$ is typically defined to be of order $\lambda^3$, while it turned out that it is  
numerically of order $\lambda^4$.} we define      
\begin{eqnarray}      
V_{ub} & = &  s_{13} e^{-i \delta_{13}} =: A \lambda^4 (\tilde{\rho} + i \tilde{\eta})     \, , 
\\      
V_{us} & = &  s_{12} (1 + {\cal O} (\lambda^8)) =: \lambda     \, , 
\\      
V_{cb} & = &  s_{23} (1 + {\cal O} (\lambda^8)) =: A \lambda^2     \, . 
\end{eqnarray}       
     
\noindent  
For the case of 4 generations one also needs the possible size, i.e. the power in $\lambda$      
of the new CKM matrix elements.   
Using the results of the previous section we find:      
\begin{align}            
|V_{ub'}|   &   \leq   0.0535 \approx 1.05 \lambda^2                        &  (1.05 \lambda^2) \, , \nn \\  
|V_{cb'}|   &   \leq   0.123  \approx 0.54 \lambda^1 \approx 2.38 \lambda^2 &  (2.8  \lambda^2) \, , \nn \\      
|V_{tb'} |  &   \leq   0.35  \approx  1.55 \lambda^1                        &  (3.0 \lambda^1)  \, . \nn      
\end{align}    
In brackets we show the results of our previous analysis in  \cite{Bobrowski:2009ng}. The biggest effect of the  
inclusion of the electroweak precision constrains was the reduction of the allowed mixing between the third and  
fourth family. The mixing between the first and the fourth family can still be considerably larger than the 
mixing between the first and the third family. This bound is still dominated by $D$-mixing. 
\begin{itemize}      
\item Defining the $V_{ub'}$ as     
      \begin{align}      
      V_{ub'}  = & s_{14} e^{-i \delta_{14}} =: \lambda^2 (x_{14} -i y_{14} ) \, , \nn       
      \end{align}  
      one obtains  
      \begin{align}  
        \Rightarrow & s_{14} = \lambda^2 \sqrt{x_{14}^2 + y_{14}^2} \, , \nn      
      \\      
        \Rightarrow & c_{14} = 1 - \lambda^4 \frac{x_{14}^2 + y_{14}^2}{2}+O\left(\lambda ^8\right)\; .    
      \end{align}      
      The parameters $x_{14}$ and $y_{14}$ are effectively   
      smaller than or equal to $1$ for all cases.     
\item Let us further define the matrix element $V_{cb'}$ via     
      \begin{align}    
      V_{cb'}&= c_{14} s_{24} e^{-i \delta_{24}}=: (x_{24} - iy_{24}) \lambda^1\;.  
     \end{align}  
      Comparison with Eq.~\eqref{eq:CKM4FP} then gives  
     \begin{align}  
       &\Rightarrow  s_{24}  e^{-i \delta_{24}} =\left(x_{24}-i y_{24}\right) \lambda +O\left(\lambda ^5\right)  \, , 
\nn \\    
&\Rightarrow c_{24}=   
\label{VcbpIsConfusing}    
1+\frac{1}{2} \left(-x_{24}^2-y_{24}^2\right) \lambda ^2+O\left(\lambda ^5\right)   \, . 
\end{align}    
 \item the last ingredient is the element $V_{tb'}$:  
 \begin{align}  
  V_{tb'}= c_{14} c_{24} s_{34} =: B \lambda  
 \end{align}  
  and therefore  
\begin{align}  
    \sin(\theta_{34})&= B \lambda +\frac{1}{2} \lambda ^3 \left(B x_{24}^2+B y_{24}^2\right)+\mathcal{O}\left(\lambda ^5\right) \, , \nn \\  
   \cos(\theta_{34})&= 1-\frac{B^2 \lambda ^2}{2}+\frac{1}{8} \lambda ^4 \left(-B^4-4 B^2 x_{24}^2-4 B^2  
   y_{24}^2\right)+\mathcal{O}\left(\lambda ^5\right) \, . 
\end{align}    
\end{itemize}

Expanding the CKM4 matrix up to and including order $\lambda^4$,    
the matrix elements take the form :  
\begin{align}   
 V_{ud}&= 1-\frac{\lambda ^2}{2}+\frac{1}{8} \lambda ^4 \left(-4 x_{14}^2-4 y_{14}^2-1\right)+O\left(\lambda  
   ^5\right) \, , & V_{us}&=\lambda \, , \nn \\   
 V_{ub}&= A ({\tilde{\rho}} -i {\tilde{\eta}}) \lambda ^4 \, , & V_{ub'}&= \left(x_{14}-i y_{14}\right) \lambda ^2 \, , \nn \\   
\end{align}   
\begin{align}   
V_{cd}=& -\lambda +\frac{1}{2} \lambda ^3 \left(x_{24}^2+y_{24}^2\right)-\left(x_{14}+i  
   y_{14}\right) \left(x_{24}-i y_{24}\right)\lambda ^4 +O\left(\lambda ^5\right) \, , \nn \\    
V_{cs}= &1+\frac{1}{2} \lambda ^2  
   \left(-x_{24}^2-y_{24}^2-1\right)+\nn \\ & + \frac{1}{8} \lambda ^4 \left(-4 A^2-2 x_{24}^2  
   \left(y_{24}^2-1\right)-x_{24}^4-y_{24}^4+2 y_{24}^2-1\right)+O\left(\lambda ^5\right) \, , 
\nn \\    
V_{cb}= & A \lambda ^2 \, , \nn \\   
V_{cb'}=& \lambda ^2 \left(x_{24}-i y_{24}\right) \, ,  
\end{align}  
\begin{align}  
V_{td}= & \lambda ^3 \left(A-B x_{14}-i B y_{14}\right)+ \left(-i A \eta -A \rho +B x_{24}+i B  
   y_{24}\right)\lambda ^4+O\left(\lambda ^5\right) \, ,\nn \\   
V_{ts}=&-A \lambda ^2-B \lambda ^3 \left(x_{24}+i  
   y_{24}\right)+\frac{1}{2} \lambda ^4 \left(A B^2-A x_{24}^2-A y_{24}^2+A-2 B x_{14}-2 i B  
   y_{14}\right)+O\left(\lambda ^5\right) \, ,\nn \\   
V_{tb}=& 1-\frac{B^2 \lambda ^2}{2}+\frac{1}{8} \lambda ^4 \left(-4  
   A^2-B^4-4 B^2 x_{24}^2-4 B^2 y_{24}^2\right)+O\left(\lambda ^5\right) \, ,\nn \\   
V_{tb'}=&B \lambda \, , 
\end{align}  
\begin{align}  
 V_{t'd}=&  \lambda ^2 \left(-x_{14}-i y_{14}\right)+\lambda ^3 \left(x_{24}+i y_{24}\right)+\nn \\ &+\frac{1}{2} \lambda  
   ^4 \left(-2 A B+\left(x_{14}+i y_{14}\right) \left(B^2+x_{24}^2+y_{24}^2\right)+x_{14}+i  
   y_{14}\right)+O\left(\lambda ^5\right) \, ,\nn \\   
V_{t's}=&\lambda ^2 \left(-x_{24}-i y_{24}\right)+\lambda ^3  
   \left(A B-x_{14}-i y_{14}\right)+\frac{1}{2} \left(B^2+1\right) \lambda ^4 \left(x_{24}+i  
   y_{24}\right)+O\left(\lambda ^5\right) \, , \nn \\   
V_{t'b}=&-B \lambda -\frac{1}{2} \lambda ^3 \left(B  
   \left(x_{24}^2+y_{24}^2\right)\right)-A \lambda ^4 \left(x_{24}+i y_{24}\right)+O\left(\lambda  
   ^5\right)  \, ,\nn \\   
V_{t'b'}=& 1+\frac{1}{2} \lambda ^2 \left(-B^2-x_{24}^2-y_{24}^2\right)+\nn \\ & + \frac{1}{8} \lambda ^4  
   \left(-B^4-2 B^2 x_{24}^2-2 B^2 y_{24}^2-2 x_{24}^2 y_{24}^2-x_{24}^4-4 x_{14}^2-y_{24}^4-4  
   y_{14}^2\right)+O\left(\lambda ^5\right) \, .  
\end{align}     
Naturally, this expansion cannot take into account correlations among the  
various CKM elements. However, the expansion is quite useful if one wants a  
rough estimate of the maximal size of a product of CKM elements determining 
the impact of the fourth generation on a certain process.

\section{Conclusion}      
\setcounter{equation}{0}

We have investigated the experimentally allowed parameter range for a hypothetical 
4$\times$4 quark mixing matrix. Therefore we extended our previous study \cite{Bobrowski:2009ng}  
of the CKM like mixing constraints of a fourth generation of quarks. 
\\   
Besides the tree-level determinations  of the 3$\times$3 CKM elements we also included 
the angle $\gamma$ of the unitarity triangle, which turned out to be a rather severe 
bound for the phases of $V_{CKM4}$. 
Next we included the electroweak $S$, $T$, $U$ parameters. 
Here we reproduced some of the results from \cite{Chanowitz:2009mz}, in particular we also excluded 
the three examples of very large mixing presented in \cite{Bobrowski:2009ng}. In this paper we have included 
for the first time the full CKM dependence of the $T$ and the $U$ parameter. 
Doing so, we found that a mass degenerate fourth family of
quarks is not excluded in that respect we differ e.g. from
\cite{Kribs:2007nz, Amsler:2008zzb}.
While degenerate quark masses can also arise if the lepton masses are
adjusted accordingly, see Sect.~\ref{ewsect} and \cite{Erler:2010sk},
including the full CKM dependence allows for a greater 'flexibility' in
the parameter space: e.g.~only then a simultaneous separate degeneracy of
leptons and quarks of the fourth generation would not be excluded ---
this may be of interest if one wants to invoke a symmetry to motivate
the tiny mass splittings. 
In addition we found also that large values  
of the parameter $U$ are not excluded a priori; only after applying 
the $T$ parameter constraint we are left with small values of $U$. 
\\ 
Concerning the FCNC constraints we studied $K$-, $D$-, $B_d$-, $B_s$-mixing and the decay $b \to s \gamma$. 
In contrast to \cite{Bobrowski:2009ng} we also included bounds to the rare decay $B_s \to \mu^+ \mu^-$ 
and we improved our treatment of the QCD corrections to $b \to s \gamma$. It turned out that the naive 
bound for  $b \to s \gamma$ used  in \cite{Bobrowski:2009ng} was too restrictive. 
\\ 
Performing a scan over the whole parameter space of the SM4  we found that typically small mixing with the fourth family  
is favored, but still some sizeable deviations from the SM3 results are not yet excluded. We demonstrated  
explicitly that e.g. effects of ${\cal O} (100 \%)$ in $B_s$ mixing are not excluded, yet. 
Concerning CP-violation in $B_s$ mixing, we could have an almost arbitrarily large phase $\Phi_s$ without 
violating the tree-level constraints. After switching on the $T$ constraint  
(99 $\%$ CL of \cite{Erler:2010sk}) we could exclude  
values of the ${\cal O} (- 50^\circ)$ for the weak mixing phase  $\Phi_s$,  
while values of  ${\cal O} (- 20^\circ)$ can easily be obtained - this is still about two  orders of  
magnitudes larger than the SM3 prediction  $\Phi_s = 0.24^\circ \pm 0.08^\circ$ \cite{Lenz:2006hd}. 
In that respect we differ slightly from \cite{Hou:2006mx,Soni:2010xh,Buras:2010pi,Hou:2010mm}. 
If the real value of $\Phi_s$ was $-50 \%$, this could not be achieved by an extension of 
the SM3, which consists only of an additional 
fermion family\footnote{This statement holds only if the current allowed range for the $S$ and $T$ parameters will 
not change.}.  
Here new results for the $B$-mixing observables from TeVatron and LHCb are very desireable.  
\\ 
We found a minimal possible value for $V_{tb}$ of 0.93 within the framework of the SM4,  
which can be compared with the  result from  single top production at the TeVatron  
\cite{Abazov:2009ii,Aaltonen:2009jj,Group:2009qk}: $V_{tb}^{\tt TeVatron} = 0.88 \pm 0.07$.  
If in nature a central value of $V_{tb} = 0.88$ was realized, this could not be explained by the SM4 alone. 
Here also more precise experimental data are desireable. 
\\ 
In general we found a delicate interplay of electroweak and flavor observables,   
which strongly suggests that a separate treatment of the two  sectors is not feasible.   
\\ 
In our opinion the next steps to determine the allowed parameter space of the SM4 consist of  
i) performing a combined electroweak and CKM fit, ii) including lepton mixing and  
iii) including even more precision observables like e.g. $R_b$ or $B \to K^* ll$. 

\section*{Acknowledgements}      
       
We thank Heiko Lacker, Jens Erler and  M. Vysotsky 
for clarifying discussions and Johann Riedl for providing us the computer code from our previous project. 
Moreover,  we thank Thorsten Feldmann for clarifying discussions concerning Ref. \cite{Buras:2010pi}.
This work was supported in part by the DFG Sonderforschungsbereich/Transregio 9
``Computergest\"utzte Theoretische Teilchenphysik''.

\end{document}